%% file: EstSpatialCorr_accepted.tex
\documentclass[a4paper,twocolumn,superscriptaddress,11pt,accepted=2018-08-28]{quantumarticle}
\usepackage{amssymb}
\usepackage[pdftex]{graphicx}	
	\DeclareGraphicsExtensions{.pdf,.png,.jpg}
\usepackage[applemac]{inputenc}
\usepackage{amsmath,amssymb,amsthm}
\usepackage{mathbbol,bbm}
\usepackage{graphicx,float}
\usepackage[final]{showkeys}
\usepackage{bm,dsfont}
\usepackage{xcolor}
\usepackage[numbers,sort&compress]{natbib}
\usepackage{booktabs}
\usepackage{enumitem}
\usepackage[symbol]{footmisc}

\newcommand{\PSComm}[1]{\textcolor{black}{#1}}

\newcommand{\MMComm}[1]{\textcolor{black}{#1}}
\newcommand{\ARComm}[1]{\textcolor{black}{#1}}
\newcommand{\LPComm}[1]{\textcolor{black}{#1}}
\newcommand{\TMComm}[1]{\textcolor{black}{#1}}
\newcommand{\RefComm}[1]{\textcolor{black}{#1}}

\newcommand{\rmS}{\mathrm{S}}

\newcommand{\calE}{\mathcal{E}}

\newcommand{\calR}{\mathcal{R}}

\def\ket#1{\left|#1\right>}
\def\bra#1{\left<#1\right|}
\def\Tr{ {\rm{Tr }}}

\begin{document}
\title{Experimental quantification of spatial correlations in quantum dynamics}

\author{Lukas Postler*}
\affiliation{Institut f\"ur Experimentalphysik, Universit\"at Innsbruck, Technikerstr. 25, A-6020 Innsbruck, Austria}

\author{\'Angel Rivas*}
\affiliation{Departamento de F\'isica Te\'orica I, Universidad Complutense, 28040 Madrid, Spain}
\affiliation{CCS-Center for Computational Simulation, Campus de Montegancedo UPM, 28660 Boadilla del Monte, Madrid, Spain}

\author{Philipp Schindler}
\affiliation{Institut f\"ur Experimentalphysik, Universit\"at Innsbruck, Technikerstr. 25, A-6020 Innsbruck, Austria}

\author{Alexander Erhard}
\affiliation{Institut f\"ur Experimentalphysik, Universit\"at Innsbruck, Technikerstr. 25, A-6020 Innsbruck, Austria}

\author{Roman Stricker}
\affiliation{Institut f\"ur Experimentalphysik, Universit\"at Innsbruck, Technikerstr. 25, A-6020 Innsbruck, Austria}

\author{Daniel Nigg}
\affiliation{Institut f\"ur Experimentalphysik, Universit\"at Innsbruck, Technikerstr. 25, A-6020 Innsbruck, Austria}

\author{Thomas Monz}
\affiliation{Institut f\"ur Experimentalphysik, Universit\"at Innsbruck, Technikerstr. 25, A-6020 Innsbruck, Austria}

\author{Rainer Blatt}
\affiliation{Institut f\"ur Experimentalphysik, Universit\"at Innsbruck, Technikerstr. 25, A-6020 Innsbruck, Austria}
\affiliation{Institut f\"ur Quantenoptik und Quanteninformation, \"Osterreichische Akademie der Wissenschaften, Otto-Hittmair-Platz 1, A-6020 Innsbruck, Austria}

\author{Markus M\"uller}
\affiliation{Department of Physics, College of Science, Swansea University, Singleton Park, Swansea - SA2 8PP, United Kingdom}

\begin{abstract}
  Correlations between different partitions of quantum systems play a
  central role in a variety of many-body quantum systems, and they
  have been studied exhaustively in experimental and theoretical
  research. Here, we investigate \textit{dynamical} correlations in
  the time evolution of multiple parts of a composite quantum
  system. A rigorous measure to quantify correlations in quantum
  dynamics based on a full tomographic reconstruction of the quantum
  process has been introduced recently~[{\'A}. Rivas et al., New
  Journal of Physics, 17(6) 062001 (2015).]. In this work, we derive a
  lower bound for this correlation measure, which does not require
  full knowledge of the quantum dynamics\TMComm{. Furthermore} we also extend the
  correlation measure to multipartite systems. We directly apply the
  developed methods to a trapped ion quantum information processor to
  experimentally characterize the correlations in quantum dynamics for
  two- and four-qubit systems. The method proposed and demonstrated in
  this work is scalable, platform-independent and applicable to other
  composite quantum systems and quantum information processing
  architectures. \PSComm{We apply the method to estimate spatial
  correlations in environmental noise processes, which are crucial
  for the performance of quantum error correction procedures.}
\end{abstract}


 
\maketitle
\footnotetext[1]{These authors contributed equally to this work.}


\input{intro}

\input{measure}

\input{lower_bound}

\input{multipartite}

\input{experimental_section}

\section{Conclusions and Outlook}


We have presented a lower bound for the measure for correlations of
quantum dynamics proposed in Ref.~\cite{NJPpaper}. We have shown how
this lower bound can be evaluated without full knowledge of the
quantum process and how it can be extended to multipartite systems. We have applied
both the full measure $\bar{I}$, which requires full tomographic information, as well as the lower bound $\bar{I}_{LB}$ to different electronic qubit encodings in a trapped ion quantum information processor.

\MMComm{Our experimentally measured values are in agreement with expected values from theoretical simulations which are based on a modelling of the \TMComm{various noise} sources for different types of qubit encoding patterns. For strings of up to four qubits our measurements confirm that the natural noise in the quantum processor characterized in this work is dominated by perfectly correlated dephasing noise. Notably, both the values of the exact measure $\bar{I}$, as obtained from quantum process tomography in the two-ion case, as well as the values of the lower bounds obtained for up to four ions, clearly reveal that the noise dynamics is no longer perfectly correlated once \TMComm{a} subset of ions is encoded in different electronic states than the others. Furthermore, the observed values for $\bar{I}$ for an asymmetric encoding also reveal the time scale on which this breaking of the perfectly correlated dephasing dynamics \TMComm{takes} effect. This quantitative information is valuable, if one is e.g.~interested in using sets of ion-qubits for the exploitation of decoherence-free subspaces.}

\PSComm{In fact, one of the most intriguing applications of the correlation
  measure $\bar{I}$ is to characterize noise processes in the context of quantum
  error correction. It should be noted that the measure itself cannot
  be directly used to assess the influence of the correlations on a
  QEC protocol, but it is able to test a specific noise model and to
  furthermore estimate the model's parameters. In particular, the measure can be used to experimentally determine the behaviour of correlated noise as a function of the distance between the constituent particles. 
 The noise model can
  then inform a microscopic model to estimate the QEC protocol's
  performance. This is due to the fact that the effect of correlations
  on error correction procedures depends strongly on the interplay of
  the actual form of the spatial correlations. The correlation measure
  quantifies the amount of correlations and thus cannot be directly
  connected to the threshold of an error correcting code without
  knowledge of the actual form of the correlations. }

\PSComm{\TMComm{In addition}, the exact correlation measure (and also the lower
  bound) can be used to certify that the amount of spatial
  correlations is reduced in qubit pairs at positions $\mathbf{r}_i$ and $\mathbf{r}_j$ with increasing distance according to certain scalings: \TMComm{Observing the way it decreases, e.g. as a powerlaw $\propto |r_i -r_j|^{-\alpha}$ with a sufficiently large exponent $\alpha$, allows one then to establish a connection to quantum error correction protocols. Such decay behaviour enters error correction protocols as a necessary condition for the provable existence of a regime in which the error correcting codes allow for fault-tolerant quantum computing \cite{Clemens-2004,Klesse-Frank-PRL-2005, Aharonov-2006, Novais-2006, Novais-2007, Preskill-2012,Novais-2013}}}. 
  
\PSComm{Complementary, the lower bound is useful to detect unexpected
  correlations in systems that do not allow full process
  tomography. It should
  be noted that the lower bound can be estimated using data that is
  taken routinely during system tune-up, such as Ramsey experiments,
  in many different physical quantum information processing
  architectures. Based on prior knowledge about the experimental system, one can also design the measured observable to be
  sensitive to a certain kind of errors.}

\PSComm{The presented methods can also be applied to characterize the
  noise environment in precision measurements. For example in
  \cite{Ruster}, two spatially separated ions are used for dc
  magnetometry. The authors also investigated the coherence time of
  the singlet state as a function of the ion distance. For short
  distances ($\approx4\,\mu$m) the ions see almost perfectly
  correlated noise resulting in no measurable decay (an almost perfect
  DFS). For larger ion separation ($\approx 6$ mm), the coherence time
  is reduced by at least one order of magnitude, indicating only
  partially correlated noise. The spatial correlation measure could be
  readily applied to this experiment to characterize the correlations
  in the noise as a function of ion distance. The distance of the
  qubits in the present work is in the order of $5 \mu$m and thus we
  expect and observe completely correlated noise.}


\section*{Acknowledgments}
We gratefully acknowledge support by the Austrian Science Fund (FWF),
through the SFB FoQuS (FWF Project No. F4002-N16), as well as the
Institut f\"ur Quanteninformation GmbH. This research was funded by
the Office of the Director of National Intelligence (ODNI),
Intelligence Advanced Reasearch Projects Activity (IARPA), through the
Army Research Office grant W911NF-16-1-0070. All statements of fact,
opinions or conclusions contained herein are those of the authors and
should not be construed as representing the official views or policies
of IARPA, the ODNI, or the U.S. Government. We also acknowledge
support by U.S. A.R.O. through grant W911NF-14-1-0103, the Spanish
MINECO grant FIS2015-67411, and the CAM research consortium QUITEMAD+
S2013/ICE-2801.

\bibliographystyle{notitles} 
\bibliography{spatial_correlations}

\clearpage
\onecolumngrid
\appendix

\input{theory_app}


\end{document}

%% file: intro.tex

\section{Introduction}
\label{sec:intro}

Correlations play a central role in quantum physics. A wide range of
quantum effects including apparently disconnected topics, such as Bell
inequalities or quantum phase transitions, can be analyzed by
considering correlations. In most of the cases these correlations refer
to those shared between different parties of a multipartite
quantum state, i.e.~describing the statistics of a system's observables at
a given time. These correlations, both classical and quantum, have been
extensively studied, quantified, and classified (see for instance \cite{Plenio-2007,Modi,Horodecki}). 
A different kind of correlations, subject of less attention, are dynamical
correlations. These account for the fact that the dynamics of one part
of a system may not be statistically independent from the dynamics of
the other parts.  Dynamical correlations in quantum systems are the
basis of many phenomena ranging from super-radiance \cite{Dicke} over
super-decoherence \cite{14-Qubit} to sub-radiance \cite{Pillet} in
atomic gases. Furthermore, the study of dynamical correlations is of central
importance in various research areas, such as
e.g.~photosynthesis and excitation transfer dynamics \cite{Caruso,Rebentrost,Nazir,Nalbach,Olbrich,JeskeRivas},
driven-dissipative phase transitions
\cite{Diehl,Verstraete,Igor,Lee,Schindler} and quantum metrology
\cite{Jeske2}. 

\PSComm{In the context of quantum information, the treatment of
  spatial correlations is highly important, but usually limited to the
  extreme cases of either completely uncorrelated noise with an
  independent noise source for each qubit, or completely correlated
  noise modeled by a single noise source with equal strength on all
  qubits \cite{Lidar-Brun-book}. One consequence of the latter type of
  correlations are decoherence-free subspaces \cite{Zanardi, Lidar1,
    Lidar2, Wineland, Haeffner}, which can be exploited in quantum
  information processing to extend the storage time of quantum states
  in noisy systems.  Understanding and quantifying these dynamical
  correlations is highly relevant for the performance of quantum error
  correction protocols~\cite{Lidar-Brun-book}, as correlated errors
  can undermine the fault-tolerant operation of quantum error
  correcting procedures \cite{preskill-review}.} \TMComm{Here, theory studies focusing on \textit{spatially} correlated noise \cite{Clemens-2004,Klesse-Frank-PRL-2005, Aharonov-2006, Novais-2006, Novais-2007, Preskill-2012,Novais-2013} have shown that in particular the distance-dependence of correlated noise can be crucial as to whether or not modified versions of the threshold theorem~\cite{ShorThreshold, PreskillThreshold} do hold. It is crucial that not only the noise strength is below a critical value, but also that (unwanted) interactions between qubits decay sufficiently \RefComm{quickly} with increasing distance. In} \PSComm{order to assess whether or not these conditions are met in experimental quantum processors, 
  theoretically well-founded and practically applicable methods to
  characterize the \textit{strength} as well as the \textit{distance-dependence} of spatial correlations} \TMComm{are required}. \PSComm{Such tools become particularly important in scalable quantum information processing architectures: There, it is forseeable that the noise environment will not be fully
  correlated in processors consisting
  of multiple smaller units that are interconnected by quantum
  channels. Noisy connection channels \TMComm{may} introduce spatial
  correlations in the system's dynamics. Similar considerations apply in
  distributed quantum systems that are interconnected by flying
  qubits.  
}

Recently, we proposed and explored a measure to rigorously quantify
dynamical correlations~\cite{NJPpaper}. \PSComm{Inspired by the
  metrics that can be employed to quantify the amount of correlations
} \TMComm{within} \PSComm{quantum states, a method to quantify dynamical correlations from
  tomographic data was proposed with the following features:}
\begin{enumerate}
\item It is a normalized quantity that is zero if and only if the
	dynamics are uncorrelated. In particular, \ARComm{it is constructed in terms of an information measure} and does not rely on any assumptions or a priori knowledge of the underlying dynamics. 
\item The quantifier introduces a hierarchy of quantum dynamics by
	enforcing a partial order relation between dynamics, i.e.~a way to quantitatively compare whether one \ARComm{bipartite} dynamic is more or less correlated \ARComm{(with respect to the evolution of its parts)} than another. This partial ordering is known as a
	\textit{fundamental law of a resource theory}~\cite{Plenio-2007, Brandao, Gour, Brandao2, Veitch, Tilmann, deVicente, Gour2}. This law
	states that the amount of correlations of some given dynamics cannot
	increase by adding uncorrelated dynamics to it, i.e.~a process for the quantifier equals
	zero. This ordering property for the amount of dynamical correlations is
	analogous to the fundamental law in the resource theory of
	entanglement \cite{Plenio-2007,Horodecki}. There, it refers to the fact
	that entanglement cannot increase under application of local
	operations and classical communication (LOCC).
\item The measure establishes a rigorous theoretical
	framework that allows for a definition and the study of
	\textit{maximally correlated dynamics} and the properties that
	such dynamics need to fulfill \cite{NJPpaper}. Again, the concept of
	maximally correlated dynamics is analogous to the one of maximally entangled
	states in the resource theory of entanglement.
\end{enumerate}
Whereas \TMComm{the proposed} quantifier for dynamical correlations~\cite{NJPpaper}
allows one to study correlated dynamics in a variety of contexts, its
general applicability comes at the price of requiring full knowledge
of the quantum dynamics. Experimentally, this requires quantum process
tomography of the full system which is only feasible in small-scale
systems, and has been experimentally demonstrated for up to three qubits \cite{Monz-PhysRevLett.102.040501}, and
quickly becomes impractical \TMComm{for} quantum systems of larger
size. \TMComm{Within this work, we therefore derive a efficiently measurable lower bound of the
quantifier applicable also to larger systems.}

\TMComm{By applying the quantifier to a two-qubit trapped-ion quantum information processor the amount of
correlations is extracted from a reconstructed process matrix. Furthermore the quantifier's lower bound is determined for dynamics in systems consisting of four trapped-ion qubits.} We investigate in
detail the noise dynamics and its correlations for different physical
encodings of the qubits that lead to different correlation
characteristics. Our findings underline the importance of
experimentally informed choices of qubit encodings in the presence of spatially
correlated noise in the context of quantum computing and quantum error
correction.

\TMComm{The presented work} is structured as follows: In Sec.~\ref{sec:meas-spat-corr}
we introduce and review the correlation measure proposed in \cite{NJPpaper}. We present the
derivation of a lower bound of the measure in
Sec.~\ref{sec:lower-bound}, followed by a generalization to
multipartite systems in Sec.~\ref{sec:extens-mult-syst}. Finally, we
present an investigation of noise dynamics in a trapped-ion system in
Sec.~\ref{sec:exp}.

%% file: measure.tex
\section{Measure for Spatial Correlations in Quantum Dynamics}
\label{sec:meas-spat-corr}

\begin{figure}[tb]
	\begin{center}
		\includegraphics[width=0.95\columnwidth]{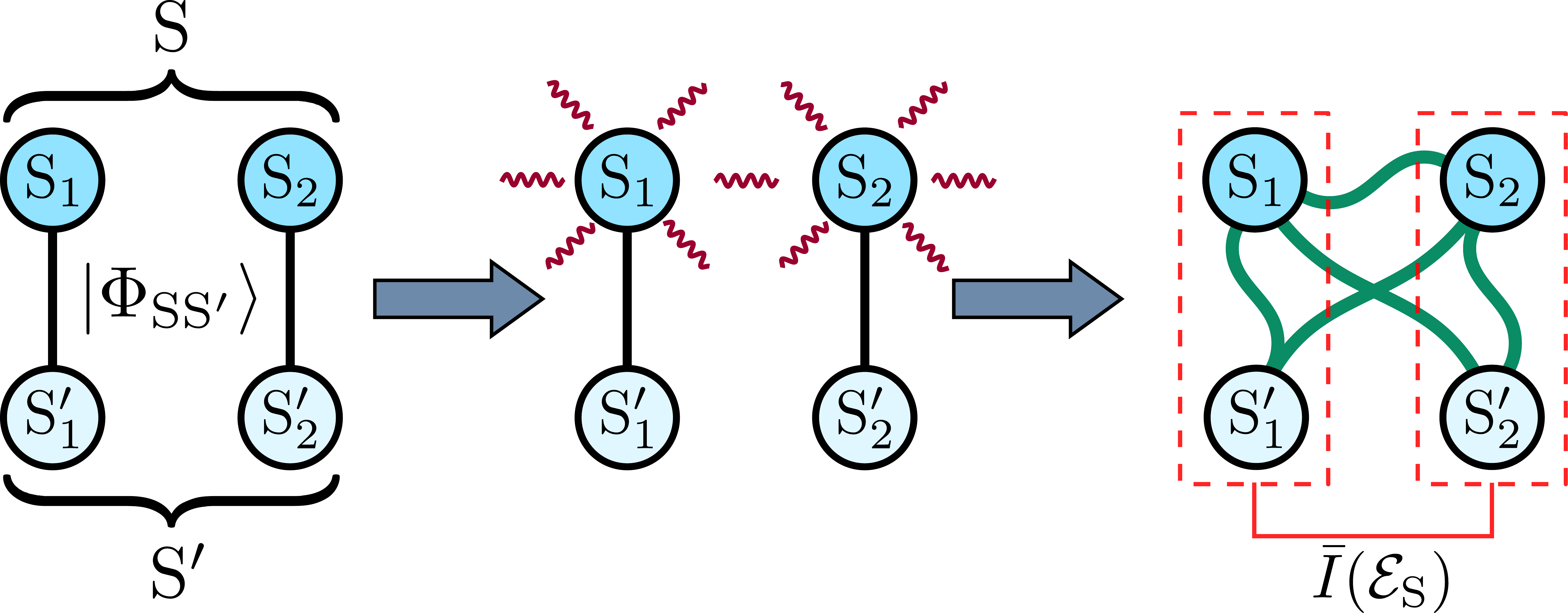}
	\end{center}
	\caption{Schematic illustration of the measurement of $\bar{I}$ in a composite system ${\rm S}$. At the beginning of the protocol subsystems $\rm{S}_1$ and $\rm{S}_2$ are maximally entangled with $\rm{S}'_1$ and $\rm{S}'_2$, respectively. The dynamics of interest is subsequently acting on $\rm{S}$ and leaves the whole system in a product state or a correlated state with respect to ${\rm S}_1 {\rm S}'_1 | {\rm S}_2 {\rm S}'_2$ depending on the amount of correlations in the dynamics.}
	\label{fig:SchemeMeasure}
\end{figure}

In the following we review the correlation measure suggested in
\cite{NJPpaper}, which is based on the Choi-Jamio{\l}kowski
isomorphism~\cite{Choi, Jamiolkowski}, providing a one-to-one mapping of the \textit{dynamics}
in a quantum system to a quantum \textit{state} of a larger system. The mathematical construction underlying the isomorphism
can be summarized as follows: consider a bipartite system
${\rm S=S_1 \otimes S_2}$, as shown in Fig.~\ref{fig:SchemeMeasure}.
Then, the idea is to initially prepare a pair of maximally entangled states $\ket{\psi^+}=\tfrac{1}{\sqrt{d}}\sum_{k=0}^{d-1}\ket{kk}$
between every part $\rm{S}_1$ and $\rm{S}_2$ with two respective ancilla systems
${\rm S}'_1$ and ${\rm S}'_2$ of the same dimension,
\begin{equation}
\ket{\Phi_{\rm SS'}}\equiv\ket{\psi^+}_{\rm S_1 S_1'}\ket{\psi^+}_{\rm S_2S_2'}=\frac{1}{d}\sum_{k=0}^{d-1}\ket{kk}\sum_{l=0}^{d-1}\ket{ll},
\end{equation}
where $d$ is the dimension of the subsystems $\rm{S}_1$ and $\rm{S}_2$, assumed to be the same for both of them for
simplicity. Then the part ${\rm S=S_1 \otimes S_2}$ of the state $\ket{\Phi_{\rm SS'}}$
evolves according to a dynamical process that is given by the map
$\mathcal{E}_{\rm S}$, and which leaves the part ${\rm S' = S'_1 \otimes S'_2}$
unperturbed. After the evolution, the final \TMComm{Choi-Jamio{\l}kowski} state
$\rho^{\rm CJ}_{\rm S}=\mathcal{E}_{\rm S}\otimes \mathds{1}_{\rm
  S'_1S'_2} (|\Phi_{\rm SS'}\rangle\langle\Phi_{\rm SS'}|)$
is a product state with respect to the bipartition ${\rm S}_1 {\rm S}'_1 | {\rm S}_2 {\rm S}'_2$,
${\rho^{\rm CJ}_{\rm S}=\rho_{\rm S_1S'_1}\otimes \rho_{\rm S_2S'_2}}$, if and
only if the dynamics are uncorrelated, 
$\mathcal{E}_{\rm S}=\mathcal{E}_{\rm S_1}\otimes\mathcal{E}_{\rm
  S_2}$.
This result follows from the one-to-one correspondence between the
map $\mathcal{E}_{\rm S}$ \TMComm{and $\rho^{\rm CJ}_{\rm S}$}. In contrast, if
the process on the composite system is correlated, then the
resulting state $\rho^{\rm CJ}_{\rm S}$ is correlated as well. A quantifier $\bar{I}$ which suitably reflects the amount of \TMComm{correlations in $\rho^{\rm CJ}_{\rm S}$ and thus in the process} $\mathcal{E}_{\rm S}$ is the quantum mutual information \cite{NC00}, given by the following definition:
\begin{widetext}
	\begin{equation}\label{Ibar}
	\bar{I}(\mathcal{E}_{\rm S}):=\frac{1}{4\log d}\left[S\left(\rho^{\rm CJ}_{\rm S}|_{\rm S_1S'_1}\right)+S\left(\rho^{\rm CJ}_{\rm S}|_{\rm S_2S'_2}\right)-S(\rho^{\rm CJ}_{\rm S})\right],
	\end{equation}
\end{widetext}
with $S(\cdot)=-\Tr[(\cdot)\log(\cdot)]$ the von Neumann entropy evaluated for $\rho^{\rm CJ}_{\rm S}$ and the reduced density operators $\rho^{\rm CJ}_{\rm S}|_{\rm S_1S'_1}:=\Tr_{\rm S_2S'_2}\left(\rho^{\rm CJ}_{\rm S}\right)$ and $\rho^{\rm CJ}_{\rm S}|_{\rm S_2S'_2}:=\Tr_{\rm S_1S'_1}\left(\rho^{\rm CJ}_{\rm S}\right)$.

\ARComm{Leaving aside the normalization factor $\tfrac{1}{4\log d}$, this quantifier can be intuitively understood as the amount of information that is needed to distinguish the actual dynamics $\mathcal{E}_{\rm S}$ from the individual dynamics of its parts $\mathcal{E}_{\rm S_1}\otimes\mathcal{E}_{\rm
  S_2}$ \cite{Schumacher}. Namely, the information that is lost when $\mathcal{E}_{\rm S_1}\otimes\mathcal{E}_{\rm
  S_2}$ is taken as an approximation of $\mathcal{E}_{\rm S}$. The normalized quantity ${\bar{I}\in[0,1]}$ quantifies this information relative to the maximum value it can take on all possible processes}.  

Furthermore, $\bar{I}$ fulfills the equation
\begin{equation}
\bar{I}\left[(\mathcal{L}_{\rm S_1}\otimes\mathcal{L}_{\rm S_2})(\mathcal{E}_{\rm S})(\mathcal{R}_{\rm S_1}\otimes\mathcal{R}_{\rm S_2})\right]\leq \bar{I}(\mathcal{E}_{\rm S})
\label{FLaw}
\end{equation}
for all local dynamical maps $\mathcal{L}_{\rm S_1}$,
$\mathcal{L}_{\rm S_2}$, $\mathcal{R}_{\rm S_1}$, and
$\mathcal{R}_{\rm S_2}$ that might act before and after the actual,
\ARComm{possibly correlated dynamics $\mathcal{E}_{\rm S}$ \cite{NJPpaper}}. Equation~\eqref{FLaw}
states that the amount of correlations of the dynamics
$\mathcal{E}_{\rm S}$ cannot increase by composition with uncorrelated
maps. In other words, if a process is a composition of a correlated
and an uncorrelated part, the amount of correlations in the
composition has to be equal or smaller than the amount of correlation that is
inherent to the correlated part. In fact, this is the fundamental law
of the resource theory of correlated dynamics \cite{NJPpaper}, where
the correlations are considered as a resource, and the operations
which do not increase the amount of this resource are uncorrelated
maps.

For clarity, we remark that the use of an ancilla system ${\rm S'}$ is underlying the mathematical construction of the isomorphism, but is not required in an experimental \ARComm{determination of $\bar{I}$}. Rather than reconstructing the Choi-Jamio{\l}kowski state $\rho^{\rm CJ}_{\rm S}$ from quantum state tomography on the enlarged system ${\rm SS'}$, one can equivalently determine \TMComm{$\rho^{\rm CJ}_{\rm S}$ by reconstructing the dynamics $\mathcal{E}_{\rm S}$ by means of quantum process tomography on the physical system S~\cite{Jezek}. Due to the Choi-Jamio{\l}kowski isomorphism in both cases the number of real parameters to determine ($4^N(4^N-1)$) is the same and \RefComm{grows exponentially with} the number of qubits.} In the following sections we therefore provide alternative strategies to estimate $\bar{I}$ avoiding full tomography.
	


%% file: lower_bound.tex

\section{Lower Bound to $\bar{I}$}
\label{sec:lower-bound}

A lower estimate for $\bar{I}$ can be obtained by performing correlation measurements on the subsystems ${\rm S_1}$ and ${\rm S_2}$.
\TMComm{Our central result is that the normalized quantity $\bar{I}(\calE_{\rm S})$ is lower bounded by
\begin{equation}\label{lowerb}
\bar{I}(\calE_{\rm S})\geq\frac{1}{8\ln d}\frac{C_{\rho'}^2(X_1,X_2)}{\| X_1 \|^2 \| X_2 \|^2},
\end{equation}
with two local quantum observables $X_1$ and $X_2$ and $C_{\rho'}(X_1,X_2)=\langle X_1\otimes X_2\rangle_{\rho'}-\langle X_1\rangle_{\rho'}\langle X_2\rangle_{\rho'}$\label{corr_meas_observables}. $\rho'=\calE_{\rm S}(\rho)$ is the evolution of an initial product state $\rho$ according to the dynamical map $\calE_{\rm S}$.
Here $\| \cdot \|$ denotes the operator norm (the absolute value of the maximum eigenvalue) 
and we have taken the logarithms inside} \ARComm{$\bar{I}(\calE_{\rm S})$ in Eq.~\eqref{Ibar} to be binary logarithms $\log_2$ (otherwise the natural logarithm $\ln d$ on the right hand side becomes multiplied by a different factor)}.

In order to prove Eq.~\eqref{lowerb}, we use the relation between the mutual information and the quantum relative entropy ${I(\rho_{\rm AB})=S(\rho_{\rm AB}\|\rho_{\rm A}\otimes\rho_{\rm B})}$ \cite{Vedral}, so that by taking the bipartition ${\rm A = S_1 S'_1}$ and ${\rm B = S_2 S'_2}$, we rewrite Eq.~\eqref{Ibar} as
\begin{equation}
	\bar{I}(\mathcal{E}_\rmS)=\frac{1}{4 \log_2 d}S\Big(\rho^{\rm CJ}_\rmS\Big\|\rho^{\rm CJ}_\rmS|_{{\rm S_1S'_1}}\otimes\rho^{\rm CJ}_\rmS|_{\rm S_2S'_2}\Big).
\end{equation}
Then the fundamental law that composition with uncorrelated dynamics does not increase the correlatedness of dynamics, as expressed in Eq.~\eqref{FLaw}, yields the inequality
\begin{equation}\label{aux1}
	\bar{I}(\mathcal{E}_\rmS)\geq\frac{1}{4 \log_2 d}S\Big(\tilde{\rho}^{\rm CJ}_\rmS\Big\|\tilde{\rho}^{\rm CJ}_\rmS|_{{\rm S_1S'_1}}\otimes\tilde{\rho}^{\rm CJ}_\rmS|_{{\rm S_2S'_2}}\Big).
\end{equation}
Here, the state
$\tilde{\rho}^{\rm CJ}_{\rm S}=\left[(\calE_{\rm
    S})(\calR_{\rm S_1}\otimes\calR_{\rm S_2})\right]\otimes\mathds{1}(|\Phi_{\rm
  SS'}\rangle\langle\Phi_{\rm SS'}|)$
is obtained by composition of the dynamics $\mathcal{E}_\rmS$ with arbitrary \text{local} maps $\calR_{\rm S_1}$ and $\calR_{\rm S_2}$, i.e.~maps that act only locally on ${\rm S_1}$ and ${\rm S_2}$, respectively. Now, monotonicity of the
relative entropy with respect to the partial trace
\cite{NC00} 
leads to the bound for the correlation measure
\begin{equation}\label{aux2}
\bar{I}(\mathcal{E}_\rmS)\geq\frac{1}{4 \log_2 d}S\Big(\tilde{\rho}^{\rm CJ}_\rmS|_{\rm S}\Big\|\tilde{\rho}^{\rm CJ}_\rmS|_{S_1}\otimes\tilde{\rho}^{\rm CJ}_\rmS|_{S_2}\Big).
\end{equation}
The trace over the subsystem ${\rm S'}$ on the Choi-Jamio{\l}kowski state yields
\begin{widetext}
	\begin{align}
	\tilde{\rho}^{\rm CJ}_\rmS|_{\rm S}:&=\Tr_{\rm S'}\left\{\left[(\calE_{\rm S})(\calR_{\rm S_1}\otimes\calR_{\rm S_2})\right]\otimes\mathds{1}(|\Phi_{\rm SS'}\rangle\langle\Phi_{\rm SS'}|)\right\}\nonumber \\ &=(\calE_{\rm S})(\calR_{\rm S_1}\otimes\calR_{\rm S_2})\left(\frac{\mathbb{1}}{d^2}\right)=\calE_{\rm S}(\rho_{\rm S_1}\otimes\rho_{\rm S_2})=\rho'.
	\end{align}
\end{widetext}
where the local maps $\calR_{\rm S_1}$ and $\calR_{\rm S_2}$ are chosen such that $\rho_{\rm S_1,S_2}:=\calR_{\rm S_1,S_2}(\tfrac{\mathbb{1}}{d})$. Moreover, since $\tilde{\rho}^{\rm CJ}_\rmS|_{\rm S_1, S_2}=\Tr_{\rm S_2,S_1}[\calE_{\rm S}(\rho_{\rm S_1} \otimes\rho_{\rm S_2})]:=\rho'_{\rm S_1,S_2}$, we write Eq. \eqref{aux2} as
\begin{equation}\label{aux3}
\bar{I}(\mathcal{E}_\rmS)\geq\frac{1}{4 \log_2 d}S(\rho'\|\rho'_{\rm S_1}\otimes\rho'_{\rm S_2}).
\end{equation}
Now we use the quantum Pinsker inequality \cite{Wilde}
\begin{equation}\label{Pinsker}
S(\rho\|\sigma)\geq \frac{1}{2\ln2}\|\rho-\sigma\|_1^2,
\end{equation}
where the relative entropy is measured in bits, $\rho$ and $\sigma$ are density matrices and the trace norm is \RefComm{$\|A\|_1=\Tr\sqrt{A^\dagger A}$}. Thus, we obtain
\begin{eqnarray}\label{aux5}
\bar{I}(\mathcal{E}_\rmS) & \geq\frac{1}{8 (\log_2 d)(\ln{2})}\|\rho'-\rho'_{\rm S_1}\otimes\rho'_{\rm S_2}\|_1^2 \nonumber \\
& =\frac{1}{8 \ln{d}}\|\rho'-\rho'_{\rm S_1}\otimes\rho'_{\rm S_2}\|_1^2.
\end{eqnarray}
Finally, by considering two arbitrary observables on ${\rm S_1}$ and ${\rm S_2}$, $X_1$ and $X_2$ respectively, the inequality $\|A\|_1\geq \frac{\Tr(AB)}{\|B\|}$ implies:
\begin{align}
\bar{I}(\mathcal{E}_\rmS)&\geq\frac{1}{8 \ln{d}}\left\{\frac{\Tr[X_1\otimes X_2(\rho'-\rho'_{\rm S_1}\otimes\rho'_{\rm S_2})]}{\|X_1\otimes X_2\|}\right\}^2\nonumber\\
&=\frac{1}{8\ln d}\frac{C_{\rho'}^2(X_1,X_2)}{\| X_1 \|^2 \| X_2 \|^2}.
\end{align}

\begin{figure}[tb]
	\begin{center}
		\includegraphics[width=0.95\columnwidth]{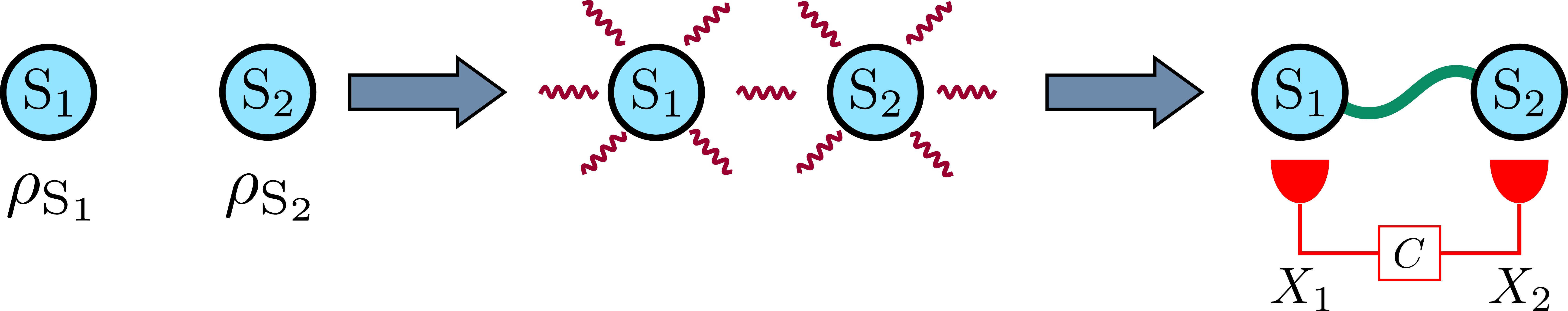}
	\end{center}
	\caption{Schematic illustration of the measurement of the lower bound
		of $\bar{I}$. The system is prepared in a separable state
		$\rho=\rho_{\rm S_1}\otimes\rho_{\rm S_2}$ and correlations of the observables
		$X_1$ and $X_2$ are measured after the evolution.}
	\label{fig:SchemeLower}
\end{figure}

\RefComm{Therefore the inequality from Eq.~\eqref{lowerb} is recovered. This result} is useful because it allows us to estimate a lower bound of the amount of dynamical correlation only by preparing product states and measuring correlation functions.
Specifically, the measurement protocol, also shown in Fig.~\ref{fig:SchemeLower}, is as follows:
\begin{enumerate}
	\item \textbf{State preparation}: The bipartite system is initially prepared in a product state $\rho=\rho_{\rm S_1}\otimes\rho_{\rm S_2}$.
	\item \textbf{Evolution}: The state $\rho$ evolves accordingly to the dynamical map $\mathcal{E}_{\rm S}$ to some state $\rho'=\calE_{\rm S}(\rho)$.
	\item \textbf{Measurement}: Correlation measurements of two local quantum observables $X_1$ and $X_2$ are carried out, $C_{\rho'}(X_1,X_2)=\langle X_1\otimes X_2\rangle_{\rho'}-\langle X_1\rangle_{\rho'}\langle X_2\rangle_{\rho'}$.
\end{enumerate}

Of course, the tightness of the inequality depends on the choice of $X_1$ and $X_2$, so in practice one has to make a well informed choice of observables, taking into account prior information about the system. \RefComm{Typically, one would choose observables that are orthogonal to the dominant noise operator in the system. However there is no simple and universal recipe to determine the observables giving the tightest lower bound.} In particular, note that finding non-vanishing values of the lower bound as given by Eq.~(\ref{lowerb}) allows one to reveal unexpected dynamical correlations if suitable observables have been estimated in any experiment.


%% file: multipartite.tex

\section{Extension to multipartite systems}
\label{sec:extens-mult-syst}

\begin{figure}[t!]
	\begin{center}
	\includegraphics[width=\columnwidth]{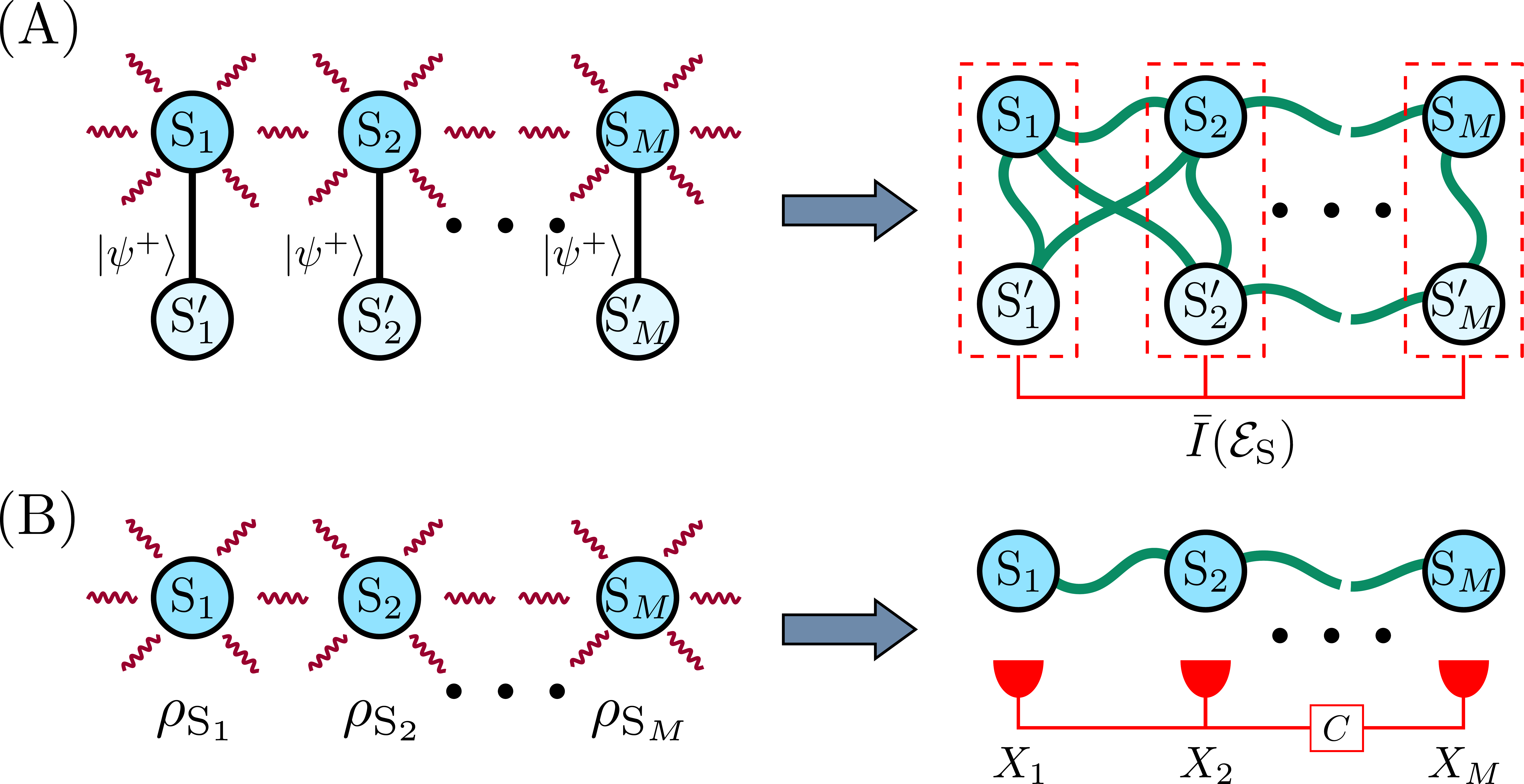}
	\end{center}
	\caption{Schematic illustration of the multipartite correlation
	  measure. (A) Choi-Jamio{\l}kowski representation of the dynamics. The
	  system is prepared in a product of maximally entangled states of $2M$ parties $\{{\rm S}_j|{\rm S}_j'\}$ and
	  the dynamics affect only the subsystems ${\rm S}_j$. If and only if the
	  dynamics are correlated, the bipartitions $\{{\rm S}_i{\rm S}'_i|{\rm S}_j{\rm S}'_j\}$ will be
	  entangled, yielding a nonzero $\bar{I}$. (B) Schematic depiction of
	  the procedure to estimate a lower bound of $\bar{I}$. There, the
	  system is prepared in a separable state
	  $\rho_{{\rm S}_1} \otimes \rho_{{\rm S}_2} \cdots \otimes \rho_{{\rm S}_M}$ and
	  correlations in the dynamics appear as \RefComm{correlations C (see page~\pageref{corr_meas_observables})} in a measurement
	  of the observables $X_j$.}
	\label{fig:CorrelationsMulti}
\end{figure}

The approach to measure and estimate bipartite correlations can be
extended to the multipartite case. In this situation, one \TMComm{has to} specify
what kind of correlations are the matter of interest. For instance, one
may be interested in the amount of correlations shared between two
parties of the system or between all parties.
Figure~\ref{fig:CorrelationsMulti} illustrates a generic situation where
correlations among all systems are investigated. 

For definiteness, suppose we consider the total amount of
correlations, i.e.~the amount of correlations shared by all parties
(other cases can be analyzed in a similar manner). In that case, if
the system $\rmS$ has $M$ parties $\rmS_1,\rmS_2,\ldots,\rmS_M$, we
introduce respective ancillary systems
$\rmS'_1,\rmS'_2,\ldots,\rmS'_M$ and prepare a collection of $M$
maximally entangled states between $\rmS_1$ and $\rmS'_1$, $\rmS_2$
and $\rmS'_2$, etc. [see Fig.~\ref{fig:CorrelationsMulti}(A)]. The dynamics are then applied on the system $\rmS$
we want to study. The amount of total normalized correlations
in the dynamics can be assessed by 
\begin{widetext}
	\begin{align}\label{IbarMulti}
		\bar{I}(\mathcal{E}_\rmS)&:=\frac{1}{2M \log d}S\Big(\rho^{\rm CJ}_\rmS\Big\|\rho^{\rm CJ}_\rmS|_{\rmS_1\rmS'_1}\otimes\ldots\otimes\rho^{\rm CJ}_\rmS|_{\rmS_M\rmS_M'}\Big)\nonumber\\
		&:=\frac{1}{2M \log d}\left\{\left[\sum_{i=1}^M S\left(\rho^{\rm CJ}_\rmS|_{\rmS_i\rmS'_i}\right)\right]-S\left(\rho^{\rm CJ}_\rmS\right)\right\},
	\end{align}
\end{widetext}
where  $\rho^{\rm CJ}_\rmS|_{\rmS_i\rmS'_i}=\Tr_{\{\forall \rmS_{j\neq i}\rmS'_{j\neq i}\}}(\rho^{\rm CJ}_\rmS)$.

The lower bound for the multipartite setting can be applied as shown
in Fig.~\ref{fig:CorrelationsMulti}(B), by measuring correlations.
Mathematically the same steps as in the bipartite case
[see Eq.~\eqref{lowerb}] can be applied, resulting in
\begin{equation}\label{lowerbMulti}
\bar{I}(\mathcal{E})\geq\frac{1}{4M\ln d}\frac{C_{\rho'}^2(X_1,\ldots,X_M)}{\| X_1 \|^2\ldots \|X_M \|^2}.
\end{equation}
Here, $\rho'$ is the joint state after the evolution of an initial product state, $X_1,\ldots,X_M$ are local observables for the parties $\rmS_1,\ldots,\rmS_M$, respectively, and
\begin{widetext}
	\begin{equation}
		C_{\rho'}(X_1,\ldots,X_M)=\langle X_1\ldots X_M \rangle_{\rho'} - \langle X_1\rangle_{\rho'}\ldots\langle X_M\rangle_{\rho'}.
	\end{equation}
\end{widetext}

This multipartite bound makes investigating correlation dynamics
accessible in systems that are too large for full quantum process
tomography\TMComm{, as the number of measurements increases only linearly compared to an exponential scaling for full quantum process tomography.}

%% file: experimental_section.tex
\section{Application to experiments}
\label{sec:exp}

\LPComm{In the following, we study the dynamics of spatial
  correlations of noise processes in a trapped ion quantum information
  processor \cite{SchindlerNJP}.  In Sec.~\ref{sec:exp_full_protocol} we
  analyze the temporal development of the spatial correlation
  estimator $\bar{I}$ to determine the degree of spatial correlations
  in a two-qubit register. For this, we perform full quantum process
  tomography on qubit registers with varying degree of correlations
  which yields following behaviour:
\begin{enumerate}[label=\alph*)]
\item Fully correlated noise enables decoherence free subspaces \TMComm{(DFS)~\cite{RoosDFS}
  and thus the correlation quantifier increases
  with decoherence, eventually reaching a steady state for times larger than the single qubit coherence time} when all single
  qubit coherences have vanished but coherences in the DFS survive. \TMComm{For purely dephasing noise the saturation value is $\bar{I} = 0.125$ (see Appendix).}
\item Partially correlated dynamics do not feature a full DFS and thus
  also, after an initial stage of increasing spatial correlations, the
  correlations will vanish \TMComm{in the limit of infinite waiting times. 
  More precisely, the correlations will start to decrease when the individual constituents of the imperfect DFS suffer substantial dephasing.}
\item For uncorrelated dynamics the quantifier should not
  detect any statistically significant correlations.
\end{enumerate}        
In \ref{sec:exp_4_ions} we utilize the lower bound to $\bar{I}$ to
characterize dynamics in a system consisting of four qubits. Initially, we
investigate the two-body correlators as a function of qubit distance
in the register. Following that, we investigate the four-body spatial
correlations for different qubit encodings.}

\LPComm{In the experimental platform used to implement the two
  protocols} each qubit is encoded in the $4S_{1/2}$ and $3D_{5/2}$
states of a single $^{40}$Ca$^+$ ion of a string of ions trapped in a
macroscopic linear Paul trap \cite{SchindlerNJP}. Doppler cooling of the ion crystal is
performed on a short-lived cycling transition between the $4S_{1/2}$
and the $4P_{1/2}$ levels, as illustrated in
Fig.~\ref{fig:energy_levels}. The same transition is used to detect
the qubit state via the electron shelving
scheme~\cite{SchindlerNJP}. Two additional repumping lasers ensure
that the ion does not get trapped in a dark state and enable resetting
from the long-lived $3D_{5/2}$ state. \TMComm{A more detailed description of the toolset and the
experimental setup used can be found in \cite{SchindlerNJP}.}

To manipulate the state of the qubit two different laser beams are
used: A global beam effectively illuminates all ions in the chain
with equal power and allows rotations on the Bloch sphere of all
qubits simultaneously. Therefore interactions of the following form
are possible:
\begin{equation}
R_{\phi}(\theta) = e^{-i\frac{\theta}{2} S_\phi},
\end{equation}
where $S_\phi = \sum_{i=0}^{N}(\sigma_x^{(i)} \cos \phi + \sigma_y^{(i)} \sin \phi)$ with $\sigma_{x,y}^{(i)}$ being single-qubit Pauli matrices acting on qubit $i$.

To perform local operations on single qubits an addressed beam is
available. This tightly focused beam is steered along the ion chain via
an electro-optical deflector. By driving the qubit transition on
resonance or in a detuned way, two types of rotations can be realised:
\begin{equation}
\begin{split}
R_{\phi}^{(j)}(\theta) = &e^{-i\frac{\theta}{2} (\sigma_x^{(j)} \cos \phi + \sigma_y^{(j)} \sin \phi)}\\
&\mathrm{and} \\
S_{z}^{(j)}(\theta) &= e^{-i\frac{\theta}{2} \sigma_z^{(j)}}.
\end{split}
\end{equation}
With this control toolset at hand we are able \TMComm{to prepare} the qubits in the required initial state,
encode them in different Zeeman sublevels and perform quantum
process tomography.

\TMComm{The degree of noise correlations between individual qubits can be tuned by encoding
them in Zeeman states with differing magnetic field
susceptibility. In $^{40}$Ca$^+$, there exist multiple possibilities to encode a qubit
in the Zeeman levels of the $4S_{1/2}$ and $3D_{5/2}$ states as
shown in Fig.\,\ref{fig:energy_levels}. The susceptibility of the qubits to the magnetic
field ranges from -2.80\,MHz/G to +3.36\,MHz/G, which allows us to
tune not only the coherence time of the individual qubits but also the correlations between qubits, when magnetic field
fluctuations are the dominant noise source. }
Understanding the dephasing dynamics, and in particular
noise correlations, in registers containing qubits in different
encodings is essential in the context of error mitigation and quantum
error correction: this understanding will be needed to determine the
viability of an approach to build, e.g.~functional logical qubits, in
complementary approaches either based on the use of spectroscopic
decoupling of ion qubits, as compared to, e.g., shuttling-based
protocols~\cite{Bermudez2017}.

\subsection{Determining the degree of spatial correlation}
\label{sec:exp_full_protocol}
\TMComm{In the following we consider dephasing dynamics that is caused by a
magnetic field acting on a string of two
ions. The various qubit-states} have different susceptibility to magnetic field
fluctuations, given by the Land\'e g~factors $g_i$ of the involved Zeeman
substates. The phase that qubit $i$ accumulates during the time
evolution is
\[ \phi_i(t) = \int_0^t d\tau B(\tau) \mu_b g_i \] with the magnitude of the
magnetic field $B(\tau)$ and the Bohr magneton $\mu_b$. The magnetic field fluctuations are modeled by multiple random
implementations of $B(t)$. The time evolution for a single
implementation can then be expressed as:
\begin{equation}
\label{eqn:deph_suscept}
U = e^{-i\phi_1 (\sigma_1^z + g \sigma_2^z)}.
\end{equation}
with the ratio of the Land\'e factors $g = g_2/g_1$. In order to
estimate the dynamics under a dephasing decay, one needs to average
the evolution over many noise realizations with random phases. A
detailed analysis of the expected decay for qubits with different
susceptibility to magnetic fields is given in the Appendix,
Sec.~\ref{sec:2_source_dephasing}.

\begin{figure}[t]
	\begin{center}
		\includegraphics[width=0.9\columnwidth]{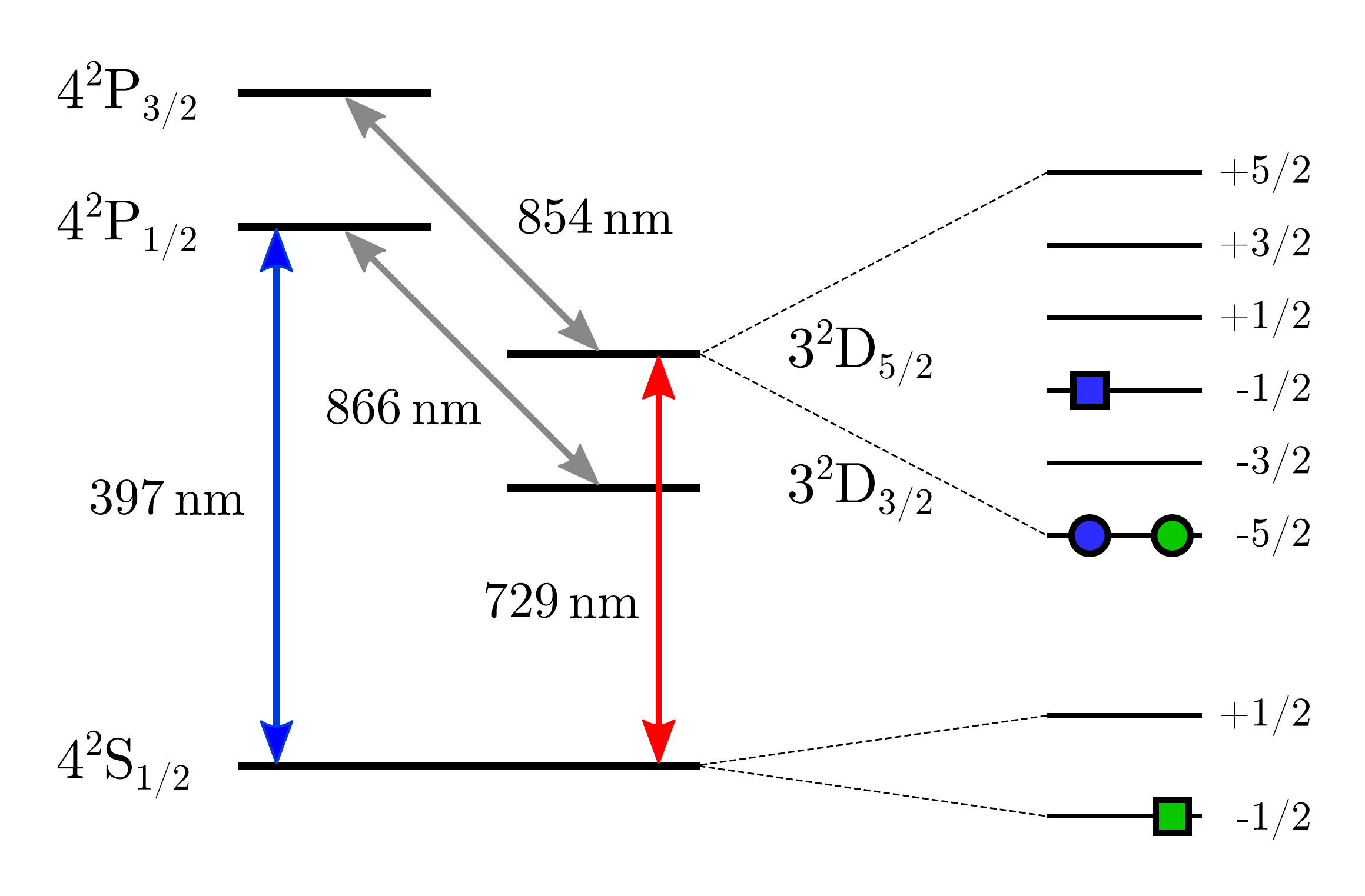}
		\caption{Level scheme of $^{40}$Ca$^{+}$. The green and
			blue squares and circles indicate different qubit
			encodings, denoted \textbf{A} and \textbf{B}, respectively. Squares are marking the qubit state
			$\ket{1}$ whereas the state $\ket{0}$ is highlighted
			with circles. The corresponding frequency shifts of
			the transitions caused by the magnetic field are -2.80\,MHz/G and +3.36\,MHz/G for the
			qubits marked with green and blue symbols
			respectively. For configuration 1 described in the enumeration in the main text for both qubits the encoding marked in green is used. The asymmetry in scenario 2 is introduced by encoding one of the qubits in in the states illustrated in blue. For the third configuration both qubits again use the encoding marked in green and the spontaneous decay from $\ket{0}$ to $\ket{1}$ is enhanced.}
		\label{fig:energy_levels}
	\end{center}
\end{figure}

In the experiment we are investigating the following qubit
configurations that implement dephasing and spontaneous decay
dynamics:
\begin{enumerate}[label=\textbf{\alph*)}]
	\item \textbf{Configuration 1:} For the realization of maximally correlated dephasing dynamics,
	both qubits are encoded in the
	$\left| 4S_{1/2}, m_S = -1/2 \right\rangle $ and
	$\left| 3D_{5/2}, m_S = -5/2 \right\rangle $ states. This encoding
	is referred to as encoding \textbf{A} hereinafter, and corresponds
	to the green markers in Fig.~\ref{fig:energy_levels}. Both qubits
	have a susceptibility to the magnetic field of -2.80\,MHz/G, leading
	to identical susceptibility coefficients ($g=1$) (see Eq.~\eqref{eqn:deph_suscept}).
	\item \textbf{Configuration 2:} To introduce an asymmetric dephasing dynamics, one qubit is
	encoded in \textbf{A} and the second is encoded in the states
	$\left| 3D_{1/2}, m_S = -1/2 \right\rangle $ and
	$\left| 3D_{5/2}, m_S = -5/2 \right\rangle $ respectively. This
	encoding is referred to as encoding \textbf{B} hereinafter, and
	corresponds to the blue markers in
	Fig.~\ref{fig:energy_levels}. Their different susceptibilities to
	magnetic field noise of -2.80\,MHz/G and +3.36\,MHz/G introduce
	unequal dephasing and therefore affect correlations between the
	qubits, corresponding to the susceptibility coefficients ($g=-0.83$).
    \item \textbf{Configuration 3:} Uncorrelated dynamics can be engineered by introducing
    spontaneous decay. In this scenario, both qubits are encoded
    in Encoding \textbf{A}. A laser pulse resonant with the
    $3D_{5/2} \leftrightarrow 4P_{3/2}$ transition at 854\,nm
    shortens the effective lifetime of the exited state by
    inducing a spontaneous decay to the $4S_{1/2}, m_S = -1/2$
    level via the $3P_{3/2}, m_S = -3/2$ level. \TMComm{This pump process implements an
    uncorrelated noise process that can be modeled as spontaneous decay.} The effective lifetime depends on
    the laser power and is in our case set to be
    $T_{spont} = 7(1)\,\mu$s.
\end{enumerate}

\begin{figure}[t]
	\begin{center}
                \includegraphics[width=0.8\columnwidth]{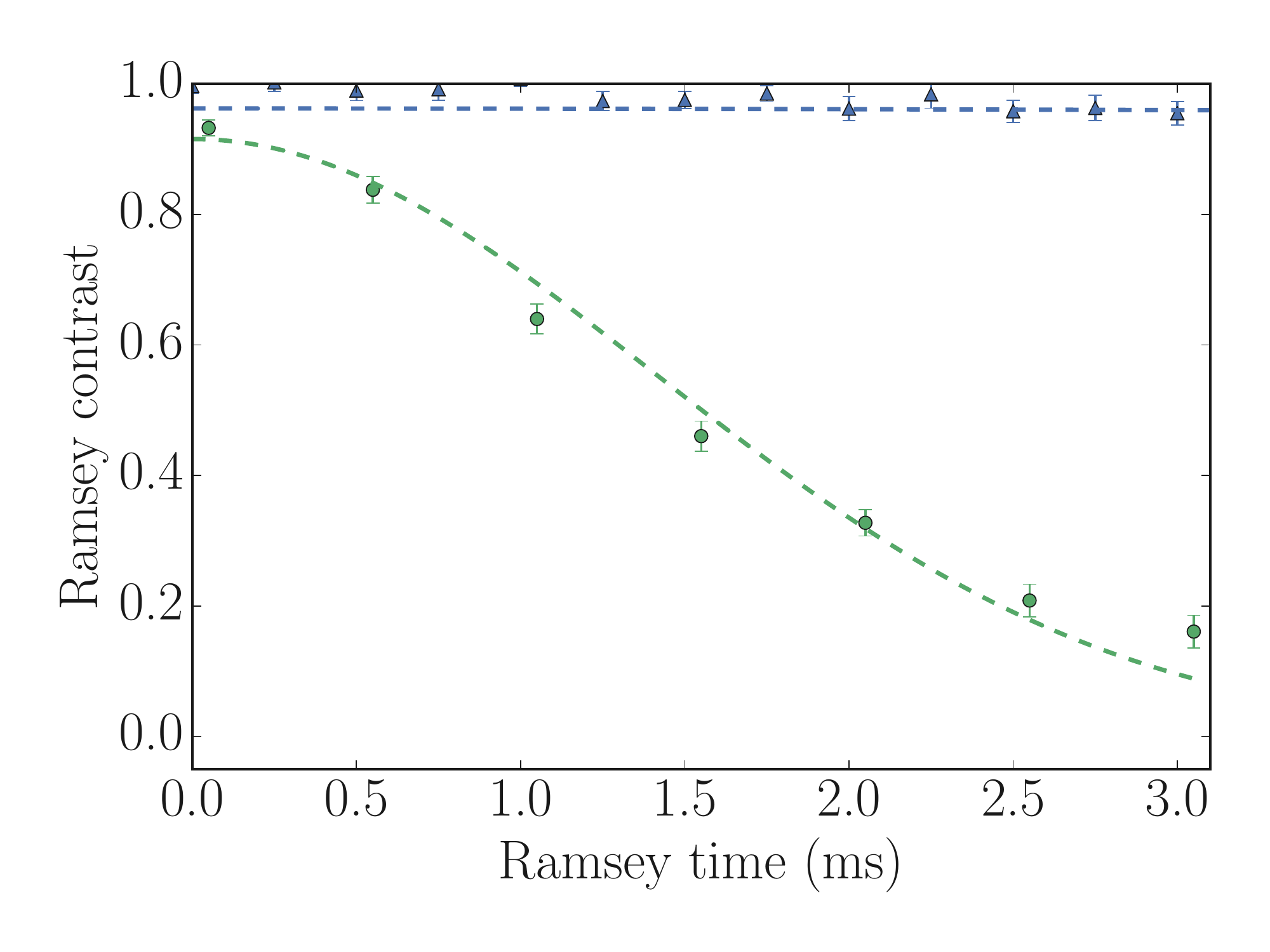}
		\caption{Coherence decay of the qubit in encoding
                  \textrm{A} without (blue triangles) and with (green
                  circles) additional magnetic field noise. }
		\label{fig:coherence}
	\end{center}
\end{figure}

\TMComm{The system size of two qubits allows us to perform full process
tomography to estimate the
correlation measure $\bar{I}$ (see Eq.~(\ref{Ibar})). In our platform for full
process tomography in an $N$ qubit system, $12^{N}$ measurement
settings, each providing $2^{N} - 1$ measurements, are required. The}
amplitude of the magnetic field fluctuations is non-stationary as it
depends on the entire laboratory environment which cannot be
controlled accurately. We engineer a stationary magnetic field
noise as the dominating noise source (a situation
where laser and magnetic field noise have to be taken into account is
described in the Appendix).  Thus we can
\TMComm{control the single qubit} coherence time as shown in Fig.~\ref{fig:coherence}. The stationary magnetic field noise is engineered by applying a white-noise current to the coils that generate the
magnetic field at the ions' positions.  We set the noise amplitude
such that the coherence time of the qubit encoded in
$\left| 4S_{1/2}, m_S = -1/2 \right\rangle $ and
$\left| 3D_{5/2}, m_S = -5/2 \right\rangle $ is reduced from
$59(3)\,$ms to $1.98(7)\,$ms. \TMComm{The increase of magnetic field noise by a factor of $\approx\,30$ ensures that laser phase noise is negligible.}
From the measured
data, a process matrix is reconstructed using an iterative Maximum
Likelihood method (see Ref.~\cite{Jezek}) that ensures trace
preservation and positivity of the process matrix.

\begin{figure}[t]
	\begin{center}
          \includegraphics[width=0.85\columnwidth]{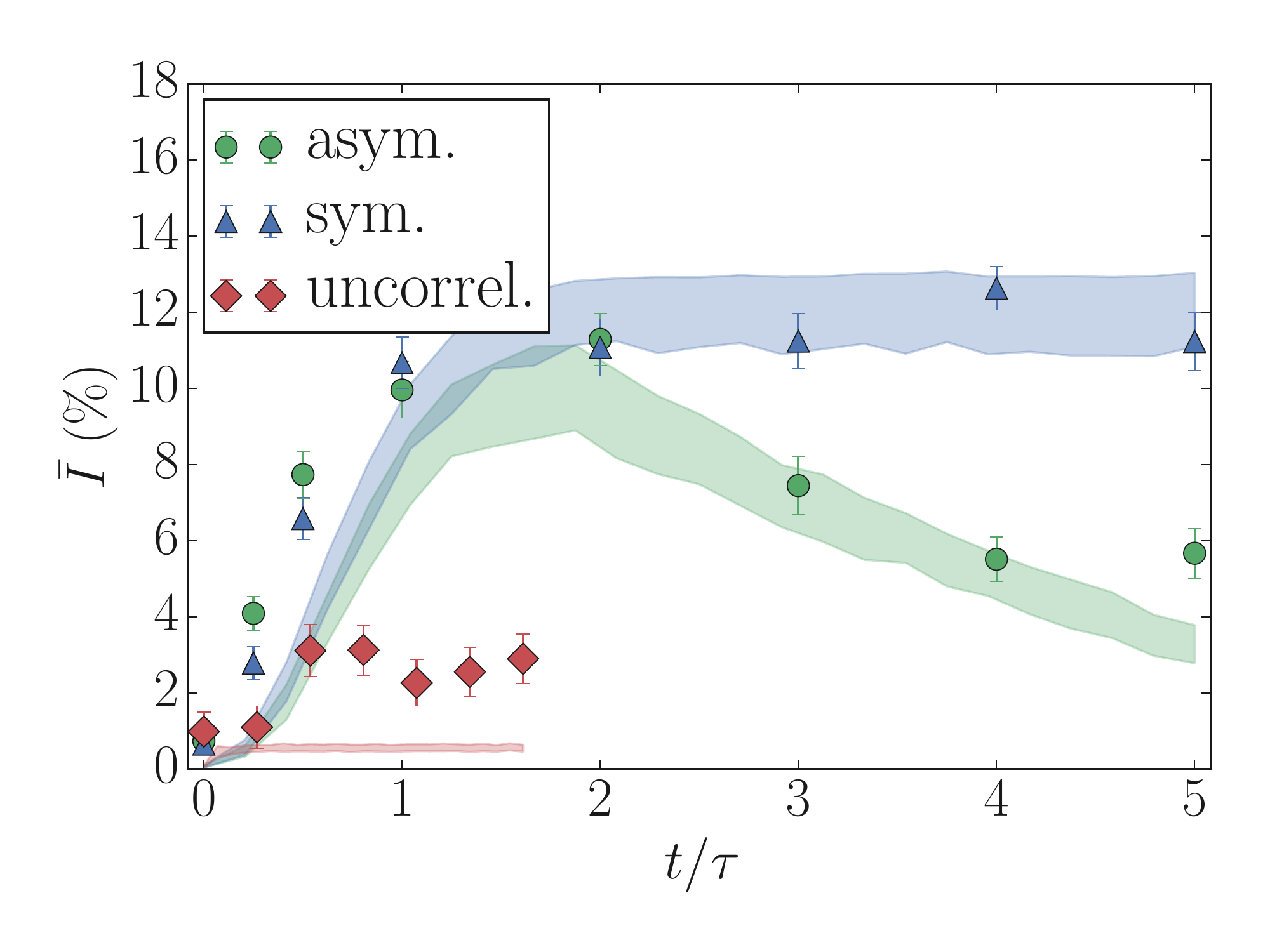}
		\caption{Dynamics of the spatial correlation quantifier $\bar{I}$ for different qubit encodings. Three cases are depicted: Both qubits encoded in $\left| 4S_{1/2}, m_S = -1/2 \right\rangle $ $\leftrightarrow$ $\left| 3D_{5/2}, m_S = -5/2 \right\rangle $ (green triangles), one qubit encoded in $\left| 4S_{1/2}, m_S = -1/2 \right\rangle $ $\leftrightarrow$ $\left| 3D_{5/2}, m_S = -5/2 \right\rangle $ and $\left| 3D_{1/2}, m_S = -1/2 \right\rangle $ $\leftrightarrow$ $\left| 3D_{5/2}, m_S = -5/2 \right\rangle$ (blue circles) and both qubits subject to uncorrelated dynamics via spontaneous decay (red diamonds). The horizontal axis is normalized to the coherence time for the first two cases and to the decay time for the third case.
		\RefComm{Results from a Monte Carlo based simulation with 500 samples are depicted with shaded areas in the corresponding color.}}
		\label{fig:exp_corr_2_ions}
	\end{center}
\end{figure}

The estimated quantifier for spatial correlations as defined in Eq.~(\ref{Ibar}) $\bar{I}$ is shown in Fig.~\ref{fig:exp_corr_2_ions} \TMComm{for the decoherence processes of} the different configurations described above. These processes are described by an exponential decay and show different timescales. To compare the data from the different configurations \TMComm{we express the free evolution time in units of the respective decay times $\tau$.} The temporal development of $\bar{I}$ is studied for evolution times of up to \TMComm{5 times the decoherence time for configurations 1 and 2 and up to 1.6 times the lifetime for configuration 3}, as the differences in the dynamics of different correlation strength are most pronounced on those timescales.

It can be seen \TMComm{in Fig.~\ref{fig:exp_corr_2_ions}} that \textbf{the symmetric configuration}
(Configuration~1), depicted with blue triangles and labeled with
\textit{"sym."}, shows the highest degree of correlations that reaches
a steady state for long evolution times.  The correlations converge to
a saturation value of 11.2(8)\,\%, which is in agreement with the
theoretical value of 12.5\,\% (as expected in the limit of perfectly
correlated dephasing) within 2 standard
deviations (see Appendix).

Measurements using \textbf{the asymmetric configuration}
(Configuration~2), depicted with green circles and labeled with
\textit{"asym."}, show similar dynamics to the symmetric setting for 
\TMComm{times up to twice the coherence time}. For longer evolution times, however, a significant decrease in
correlations is observed as no DFS is available in the system.

The third investigated scenario (Configuration~3) implementing engineered
\textbf{uncorrelated} dynamics by adding spontaneous decay, is depicted with
red diamonds.  The correlations do not exceed a value of 3.1(6)\,\% in
this case. This is significantly lower than the maximum of $\bar{I}$
for fully and partially correlated dephasing dynamics.

The blue shaded area in the figure shows simulated results
where random phase fluctuations are acting on a two-qubit
system. From the resulting output state of the simulation we generate data 
including projection noise and the
same analysis as for the experimental data is performed. To
simulate the asymmetric configuration the applied random phase
fluctuations are acting on the qubit weighted according to the
different susceptibilities to the magnetic field. For the simulation
of uncorrelated spontaneous emission, instead of phase fluctuations, 
probabilistic, uncorrelated quantum jump trajectories of the
individual qubits are simulated. A more detailed description of the
simulation can be found \TMComm{in the Appendix}.
There is qualitative agreement between simulations and
measurements, but still there are significant \TMComm{deviations, especially
in the case of uncorrelated dynamics, of up to approximately $4\,\sigma$}. We assume that this
overestimation of the spatial correlations in the system dynamics by
the quantifier is due to miscalibration and drifts of experimental
parameters. \RefComm{For instance a mismatch between the actual and the calibrated Rabi frequency would lead to additional correlated errors during the process tomography. This effect is best visible for Configuration~3, where the dynamics are expected to show no correlations at all.}

\subsection{Spatial correlations in multi-qubit systems}
\label{sec:exp_4_ions}

Due to the exponential scaling of the number of measurement settings
for a single full process tomography an analysis for a four qubit
system would take about 24 hours per waiting time setting and
configuration in our system. \TMComm{Therefore, the feasible method to investigate correlations is based on the 
provided lower bound for the correlation measure as described in Sec.~\ref{sec:extens-mult-syst}.}
For these experiments, the free-evolution time is fixed to 10\,ms,
approximately corresponding to 5 times the coherence time. 
\TMComm{Inferring from the two-qubit measurements, this evolution time renders a reliable discrimination of encoding configurations, leading to differing degrees of correlations, possible.}

\begin{figure}[t]
	\begin{center}
		\includegraphics[width=0.75\columnwidth]{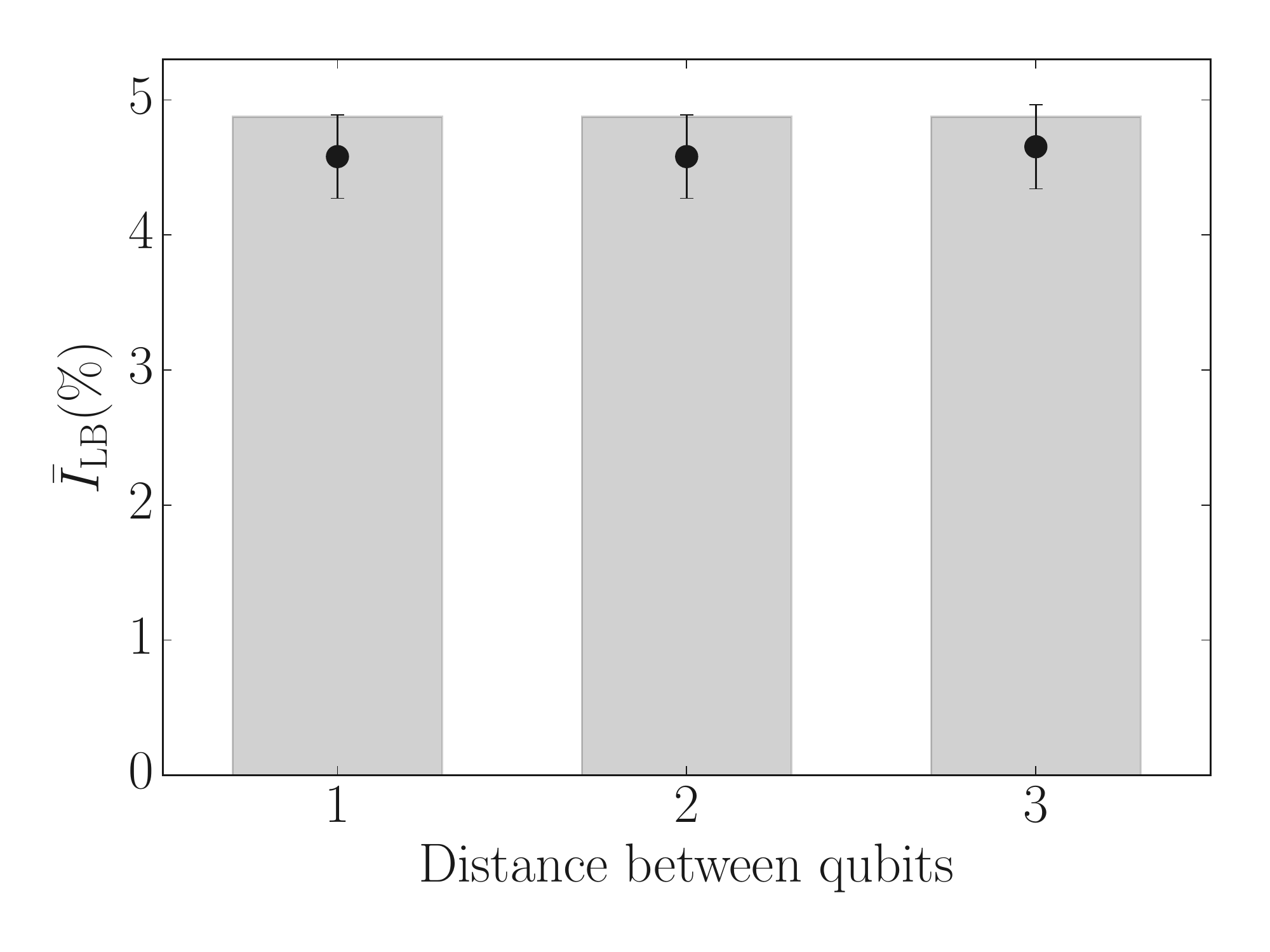}
		\caption{Lower bound for pairwise spatial correlations in a four-qubit system as a function of the distance (in terms of ion index difference in the ion string) to qubit 1, depicted with black circles. The corresponding simulations are shown with shaded bars. As all qubits have the same encoding, the correlations between qubit 1 and qubit 2, 3 and 4 respectively are the same within errorbars.}
		\label{fig:corr_2_out_of_4}
	\end{center}
\end{figure}

\PSComm{First, we investigate the distance dependence of pairwise
  spatial correlations along the register. For this we evaluate the
  lower bound for $\bar{I}$ (Eq.~(\ref{lowerbMulti})) between the
  outermost qubit and subsequently all other qubits. In a four-qubit
  system this corresponds to evaluating the observables $X_1X_i$, with \RefComm{$X_i = \bigl( \begin{smallmatrix}0 & 1\\ 1 & 0\end{smallmatrix}\bigr)_i$ acting on ion
  $i \in \{2,3,4\}$,} yielding the lower bound
  $\bar{I}_{LB}=\frac{1} {4\cdot2\ln 2}\left[\langle X_1 X_i \rangle - \langle X_1\rangle \langle X_i\rangle\right]^2$. \RefComm{By applying a global
  $\pi/2$ pulse around the x-axis of the Bloch sphere $R_0(\pi/2)$ the system is prepared in the state $\ket{+}~:=~\frac{1}{\sqrt{d}}\sum_{k=1}^d\ket{k}$, where $d = 2^4$ for four ions$^\dagger$. After the preparation the system undergoes a free evolution. With this preparation the qubits are in an eigenstate of $X_i$, the expectation values of these observables are decaying quickly due to the dominant dephasing caused by the excess magnetic field fluctuations. Therefore the lower bound of the dynamical correlations is the tightest for this choice of observables.} The data as shown
  in Fig.~\ref{fig:corr_2_out_of_4} indicate that the spatial
  correlations do not show any distance dependence. We compare the
  measured values by numerical simulations which are shown as shaded
  bars in Fig.~\ref{fig:corr_2_out_of_4} (see also Appendix). The
  measured values are close to the expected value for a fully dephased
  state under perfectly correlated noise of 4.5\% (see
  Appendix). }
\footnotetext[2]{\RefComm{In that notation the computational basis is numbered consecutively, so $\ket{1}=\ket{0000}$, $\ket{2}=\ket{0001}$, ..., $\ket{16}=\ket{1111}$.}}

\PSComm{We then analyze the spatial correlations as given by the
  four-body observable $X_1X_2X_3X_4$. We investigate three different
  configurations of qubit encodings which give rise to different
  decoherence free subspaces:}

\begin{itemize}
	\item \textbf{Configuration 1:} All four qubits encoded in Encoding \textbf{A}.
	\item \textbf{Configuration 2:} Qubits 1 and 2 encoded in Encoding \textbf{A}, qubits 3 and 4 encoded in Encoding \textbf{B}.
	\item \textbf{Configuration 3:} 
	Qubits 1, 2 and 3 encoded in Encoding \textbf{A}, qubit 4 encoded in Encoding \textbf{B}. 
\end{itemize}

\begin{figure}[t]
	\begin{center}
		\includegraphics[width=0.9\columnwidth]{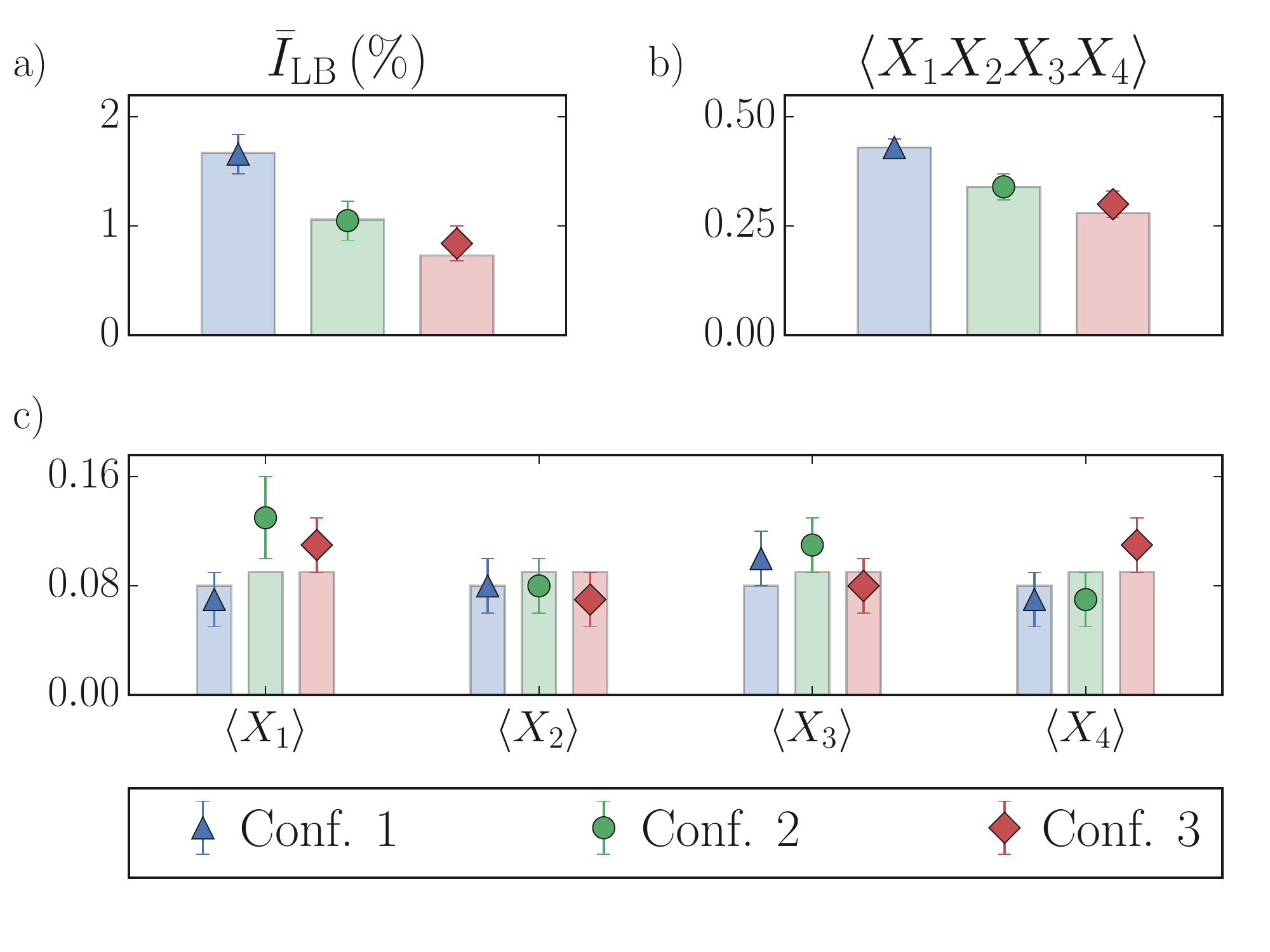}
		\caption{a) Lower bound $\bar{I}_{LB}$ for $\bar{I}$ for 3 different qubit encoding configurations described in Section~\ref{sec:exp_4_ions}. b) and c) show the underlying expectation values from which $\bar{I}_{LB}$ is calculated. The expected results from numerical simulations are depicted with shaded bars in the corresponding color.}
		\label{fig:exp_4_ions}
	\end{center}
\end{figure}

\RefComm{The preparation of the system is the same as in the measurement of the pairwise correlations, but after} the waiting time, a four qubit
state tomography measurement is performed. From the reconstructed
density matrix, we estimate the expectation values of the observables
$\langle X_1\rangle$, $\langle X_2\rangle$, $\langle X_3\rangle$,
$\langle X_4\rangle$ and $\langle X_1 X_2 X_3 X_4 \rangle$, \RefComm{where
$X_i = \bigl( \begin{smallmatrix}0 & 1\\ 1 & 0\end{smallmatrix}\bigr)_i$ is a Pauli matrix acting on qubit $i$.}
\PSComm{The experimental results for the lower bound}
\LPComm{$\bar{I}_{LB}=\frac{1}{4\cdot4\ln 2}\left[\langle X_1 X_2 X_3
    X_4 \rangle - \langle X_1\rangle \langle X_2\rangle \langle
    X_3\rangle \langle X_4\rangle \right]^2$} \PSComm{but also for the
  individual expectation values are presented in
  Fig.~\ref{fig:exp_4_ions}. In the left subplot in the top row the
  lower bound is plotted in green, blue and red for the qubit encoding
  configurations \textbf{1}, \textbf{2} and \textbf{3},
  respectively. The theoretically expected values from numerical
  simulations of the microscopic noise are depicted with shaded bars.}
These estimated correlations and single-qubit expectation values lead
to a lower bound of 1.66(18)\,\%, 1.05(18)\,\% and 0.84(16)\,\% for
the three qubit encoding patterns, respectively. These results show
agreement with simulations within 1 standard deviation. \RefComm{Note
  furthermore that in particular the lower bound value of 1.66(18)\,\%
  for the first encoding pattern in which all four qubits reside in
  the Encoding \textbf{A} (Configuration 1) is a signature of the
  almost perfectly correlated dynamics - it is close to the theory
  value predicted for the long time limit of perfectly correlated
  dephasing dynamics of 1.27\,\%. In contrast, for two qubits in Encoding \textbf{A} and two qubits in \textbf{B} (Configuration 2), the observed bound of 1.05(18)\,\% is lower than for Configuration 1 and to be compared with the theory value of 0.56\,\% predicted for the long time limit (see Appendix). For the asymmetric Configuration 3 (three qubits in Encoding \textbf{A} and one in \textbf{B}), one observes with 0.84(16)\,\% the lowest value of the lower bound. As shown in the Appendix, theory predicts the lower bound to fully vanish in the limit of even longer waiting times.}

\TMComm{Instead of performing state tomography the expectation values necessary to calculate the lower bound of the correlation quantifier could be measured directly, leading to a linear scaling of the number of measurements with the number of qubits in the worst case. In exchange for the better scaling the correlation estimation gets vulnerable to projection noise. By increasing the number of repetitions of the experiment this error source can be reduced to arbitrary low levels.}

%% file: theory_app.tex
\section{Simulations}
\label{sec:2_source_dephasing}
It is common to find more than one source of dephasing in ion trap architectures. Here we consider the case where one type of dephasing is caused by the external and global magnetic field fluctuations and the other due to the laser frequency fluctuations of the addressing laser source. In particular, with respect to the contribution from magnetic field fluctuations, it is important to note that trapped ion qubits encoded in the electronic qubit levels $S_{1/2}(m_{j}=-1/2) = |1\rangle$ $\rightarrow$ $D_{5/2}(m_{j}=-1/2) = |0\rangle$ will undergo a different dephasing dynamics than temporarily spectroscopically decoupled (``hidden'') qubits.
\TMComm{Spectroscopical decoupling has been commonly employed e. g. in the context of repetitive quantum error correction \cite{SchindlerQEC}, entangling subsets of qubits \cite{NiggColorCode} and quantum teleportation \cite{Teleportation}.}

Here, we consider \TMComm{a} string of two ions, the first of which is spectroscopically decoupled, and the second one residing in the standard qubit subspace. Thus, we effectively consider two ions with a laser field addressing the second one. The joint state of the ions $\rho$ undergoes the evolution which corresponds to accumulating random phases, 
\begin{equation}
e^{-i\phi_B (a\sigma_1^z+b\sigma_2^z)}e^{-i\phi_L\sigma_2^z},
\end{equation}
where $\phi_B$ is due to the global magnetic field fluctuations and $\phi_L$ to frequency fluctuations of the laser source addressing the second ion. The constants $a$ and $b$ depend on the specific energy levels on each ion taken into consideration. If those are the same, we have $a=b=1$. Denoting by $p_B(\phi_B)$ and $p_L(\phi_L)$ the probability distributions for $\phi_B$ and $\phi_L$, respectively, we have the noisy dynamics
\begin{equation}
\mathcal{E}(\rho)=\int d\phi_B p_B(\phi_B) \int d\phi_L p_L(\phi_L) [e^{-i\phi_B (a\sigma_1^z+b\sigma_2^z)}e^{-i\phi_L\sigma_2^z}]\rho [e^{i\phi_B (a\sigma_1^z+b\sigma_2^z)}e^{i\phi_L\sigma_2^z}].
\end{equation}
We can write
\begin{align}
&e^{-i\phi_B a\sigma_1^z}=\cos(a \phi_B)-i\sin(a \phi_B )\sigma_1^z,\\
&e^{-i(b\phi_B +\phi_L) \sigma_2^z}=\cos(b\phi_B +\phi_L)-i\sin(b\phi_B +\phi_L)\sigma_2^z,
\end{align}
so that
\begin{align}
e^{-i\phi_B a\sigma_1^z}e^{-i(b\phi_B +\phi_L) \sigma_2^z}&=\cos(a \phi_B)\cos(b\phi_B +\phi_L) \nonumber\\
&-i\sin(a \phi_B )\cos(b\phi_B +\phi_L)\sigma_1^z\nonumber \\
&-i\cos(a \phi_B)\sin(b\phi_B +\phi_L) \sigma_2^z \nonumber\\
&- \sin(a \phi_B )\sin(b\phi_B +\phi_L)\sigma_1^z\sigma_2^z,
\end{align}
and then
\begin{equation}
\mathcal{E}(\rho)=\sum_{\alpha,\beta=0,\dots3} \chi_{\alpha\beta}G_\alpha \rho G_\beta,
\end{equation}
with $G_0=\mathbb{1}\otimes \mathbb{1}$, $G_1:=\sigma_1^z$, $G_2:=\sigma_2^z$, and $G_3:=\sigma_1^z\sigma_2^z$. The coefficients $\chi_{\alpha\beta}$ form a self-adjoint matrix with the following components:
\begin{align}
\chi_{00}&=\int \int d\phi_B d\phi_L p_B(\phi_B)  p_L(\phi_L)\cos^2(a \phi_B)\cos^2(b\phi_B +\phi_L), \nonumber \\
\chi_{01}&=\frac{i}{2}\int \int d\phi_B d\phi_L p_B(\phi_B)  p_L(\phi_L)\sin(2a \phi_B)\cos^2(b\phi_B +\phi_L), \nonumber\\
\chi_{02}&=\frac{i}{2}\int \int d\phi_B d\phi_L p_B(\phi_B)  p_L(\phi_L)\cos^2(a \phi_B)\sin[2(b\phi_B +\phi_L)], \nonumber\\
\chi_{03}&=-\frac{1}{4}\int \int d\phi_B d\phi_L p_B(\phi_B)  p_L(\phi_L)\sin(2a \phi_B)\sin[2(b\phi_B +\phi_L)],\nonumber\\
\chi_{11}&=\int \int d\phi_B d\phi_L p_B(\phi_B)  p_L(\phi_L)\sin^2(a \phi_B)\cos^2(b\phi_B +\phi_L), \nonumber\\
\chi_{12}&=\frac{1}{4}\int \int d\phi_B d\phi_L p_B(\phi_B)  p_L(\phi_L)\sin(2a \phi_B)\sin[2(b\phi_B +\phi_L)],\nonumber\\
\chi_{13}&=\frac{i}{2}\int \int d\phi_B d\phi_L p_B(\phi_B)  p_L(\phi_L)\sin^2(a \phi_B)\sin[2(b\phi_B +\phi_L)],\nonumber\\
\chi_{22}&=\int \int d\phi_B d\phi_L p_B(\phi_B)  p_L(\phi_L)\cos^2(a \phi_B)\sin^2(b\phi_B +\phi_L),\nonumber\\
\chi_{23}&=\frac{i}{2}\int \int d\phi_B d\phi_L p_B(\phi_B)  p_L(\phi_L)\sin(2a \phi_B)\sin^2(b\phi_B +\phi_L),\nonumber\\
\chi_{33}&=\int \int d\phi_B d\phi_L p_B(\phi_B)  p_L(\phi_L)\sin^2(a \phi_B)\sin^2(b\phi_B +\phi_L).
\label{eqn:chi_matrix}
\end{align}

If we consider a Gaussian distribution for every random phase
\begin{equation}
f(\phi_B)=\frac{1}{\sqrt{2\pi}\sigma_B}e^{-\tfrac{\phi_B^2}{2\sigma_B^2}} \quad\text{and}\quad f(\phi_L)=\frac{1}{\sqrt{2\pi}\sigma_L}e^{-\tfrac{\phi_L^2}{2\sigma_L^2}},
\end{equation}
we obtain $\chi_{01}=\chi_{02}=\chi_{13}=\chi_{23}=0$ because of the odd parity of the integrating functions.

As we assumed to have pure dephasing dynamics the Choi-Jamio{\l}kowski state can be written as:

\begin{align*}
\rho^{\rm CJ}_{\rm S} &= \mathcal{E}_{\rm Z}\otimes \mathds{1} (|\Phi_{\rm SS'}\rangle\langle\Phi_{\rm SS'}|)\\
 &= \frac{1}{d^2} \sum^d_{k,l,m,n = 1} \alpha_{klmn}\ket{kl}\bra{mn} \otimes \ket{kl}\bra{mn}
\end{align*}

All diagonal elements are $\alpha_{klkl}=1$, and the remaining integrals in Eq.~(\ref{eqn:chi_matrix}) can analytically be performed, yielding the following components of the Choi-Jamio{\l}kowski state $\rho^{\rm CJ}_{\rm S}$:
\begin{align}
\alpha_{1112}&=\alpha_{2122}=\chi_{00}+\chi_{11}-\chi_{22}-\chi_{33} = e^{-2(b^2\sigma^2_B+\sigma_L^2)}\nonumber \\
\alpha_{1121}&=\alpha_{1222}=\chi_{00}-\chi_{11}+\chi_{22}-\chi_{33}=e^{-2a^2\sigma^2_B}\nonumber\\
\alpha_{1122}&=\chi_{00}+2\chi_{03}-\chi_{11}-2\chi_{12}-\chi_{22}+\chi_{33}=e^{-2[(a+b)^2\sigma^2_B+\sigma_L^2]}\nonumber \\
\alpha_{1221}&=\chi_{00}-2\chi_{03}-\chi_{11}+2\chi_{12}-\chi_{22}+\chi_{33}=e^{-2[(a-b)^2\sigma^2_B+\sigma_L^2]}.
\end{align}

\begin{figure*}[tb]
\begin{center}
\includegraphics[width=0.79\textwidth]{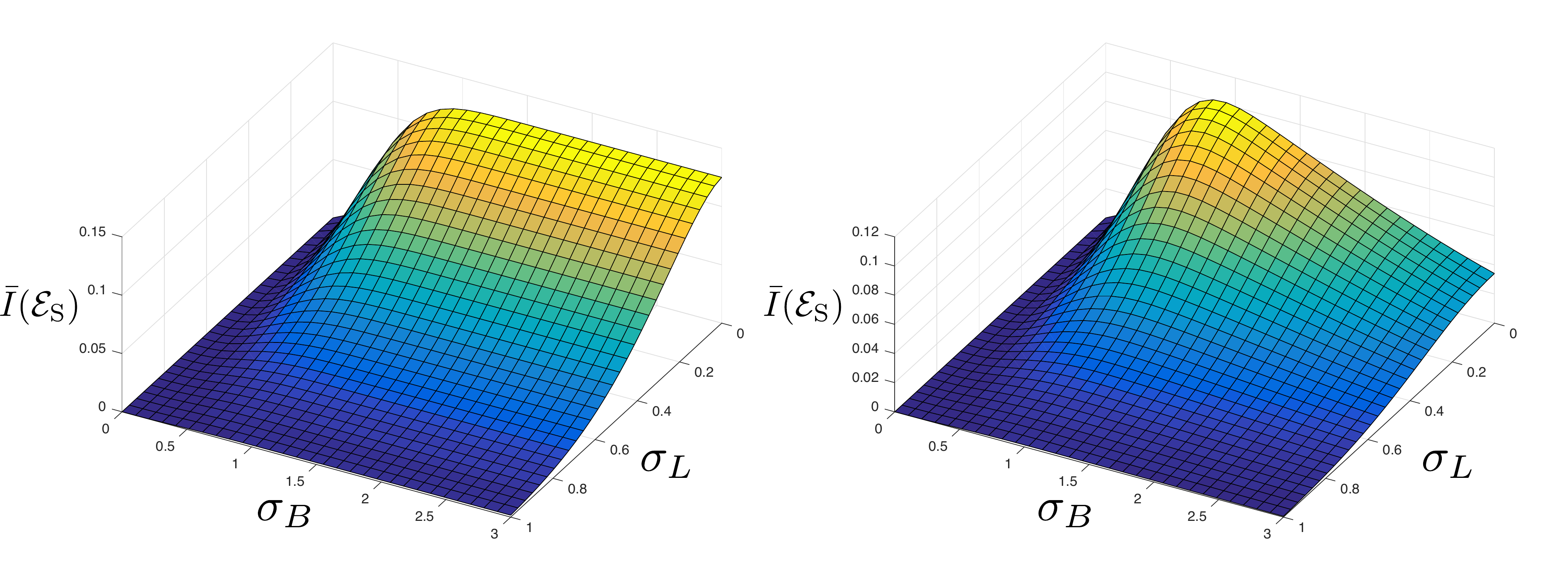}
\end{center}
\caption{Amount of correlations for $a=\pm b$ (left) and $a=1$, $b=-0.83$ (right), as a function of $\sigma_B$ and $\sigma_L$.}
\label{fig:correlationsSymmetry}
\end{figure*}

\PSComm{This allows us to compute the measure of correlation $\bar{I}$ from the CJ state. The results are shown in Fig. \ref{fig:correlationsSymmetry} for $a=\pm b$, and for $a=1$, $b=-0.83$. For $a\neq \pm b$ the amount of correlations has a maximum for some value of $\sigma_B$ and $\sigma_L=0$, and decays for large $\sigma_B$ or $\sigma_L$. In the case of $a=\pm b$, where both systems have the same susceptibility to magnetic field fluctuations, a maximum value for $\bar{I}$ of 0.125 is reached in the limit of $\sigma_B\rightarrow\infty$ and $\sigma_L=0$. For the experimental implementations from section~\ref{sec:exp} enhanced magnetic field noise, engineered by applying white current noise to coils in the ion's surrounding, was added, rendering the laser phase noise described by $\sigma_L$ negligible. Therefore, the presented experimental results correspond to a cut through these 3D~figures at $\sigma_L=0$.
Configurations~1 in section~\ref{sec:exp_full_protocol} corresponds to $a=b$ (see left part of Fig.~\ref{fig:correlationsSymmetry}). The asymptotic limit of 0.125 is in agreement with the experimental results in Fig. \ref{fig:exp_corr_2_ions}. Configuration~2 is corresponding to $\frac{b}{a} = -0.83$.
}

\LPComm{To simulate the dynamics of the build-up of space correlations in Fig.~\ref{fig:exp_corr_2_ions} a different simulation method is used:
	Random phase fluctuations are acting on both qubits, where the experimental waiting time is corresponding to the width of the phase distribution from which the samples are drawn. \TMComm{After 1000 realizations of random dephasing
	dynamics, the resulting density matrix is used to generate simulated
	measurement results by calculating expectation values for all combinations of Pauli operators acting on the two
	qubits, which correspond to the probabilities to measure a certain
	output state. From this set of probabilities measurement results are
	generated using a multinomial distribution.
	We estimate the error arising from this statistical error (projection noise) by performing 100 realizations of the simulation.
	The methods used to simulate results for the asymmetric configuration are the same, apart from the fact} that the phase fluctuation of random
	strength is applied to qubit 1 directly and multiplied by the factor
	to include the different susceptibilities to the magnetic field due
	to the different Land\'e g~factors of the included states before
	acting on qubit 2. For the simulation of the uncorrelated case a
	slightly different procedure is used: Instead of random dephasing, an
	independent decay to the ground state of the two qubits is applied
	to reflect the uncorrelated dynamics due to enhanced spontaneous
	decay.} 
	\vspace{5mm}\\
\textbf{Analytical expectations for the long-time limit under dephasing dynamics}\newline\\
\textit{4-qubit correlations - } It is instructive to consider also the long-time dynamics in the case of perfectly correlated dephasing dynamics, and the situation in which one is interested in obtaining the lower bound of the correlation measure. \RefComm{Let us first focus on \textbf{Configuration 1}, in which all four qubits are encoded in Encoding A}. The initial product state of the four qubits, $\ket{\psi} = \ket{+}^{\otimes 4}$, can be written as 
\begin{align}
\ket{\psi} = \frac{1}{4} (\ket{\psi_0} + \ket{\psi_4}) + \frac{1}{2} (\ket{\psi_1} + \ket{\psi_3}) + \sqrt{\frac{3}{8}} \ket{\psi_2}
\end{align}
with the Dicke states $\ket{\psi_j}$ of $j=0$ up to $j=4$ excitations, 
\begin{align}
\ket{\psi_0} & = \ket{0000}, \nonumber\\ 
\ket{\psi_1} & = \frac{1}{2} (\ket{1000} + \ket{0100} + \ket{0010} + \ket{0001}),\nonumber\\ 
\ket{\psi_2} & = \frac{1}{\sqrt{6}} (\ket{0011} + \ket{0101} + \ket{0110} + \ket{1001} + \ket{1010} + \ket{1100}), \nonumber\\ 
\ket{\psi_3} & = \frac{1}{2} (\ket{0111} + \ket{1011} + \ket{1101} + \ket{1110}),\nonumber\\ 
\ket{\psi_4} & = \ket{1111}. \nonumber
\end{align}
Under spatially perfectly correlated dephasing, the initial state $\ket{\psi}$ evolves for times much longer than the single-qubit coherence time, but still shorter than the life-time of the metastable qubit state, into 
\begin{align}
\ket{\psi} \bra{\psi} \stackrel{t \rightarrow \infty}{\longrightarrow} \frac{1}{16} (\ket{\psi_0} \bra{\psi_0} + \ket{\psi_4} \bra{\psi_4} ) + \frac{1}{4} (\ket{\psi_1} \bra{\psi_1} + \ket{\psi_3} \bra{\psi_3} ) + \frac{3}{8} \ket{\psi_2} \bra{\psi_2},
\end{align}
i.e.~the subspaces of a fixed excitation number $j$ are decoherence-free, so that the initial coherences between basis states within one and the same excitation number $j$ subspace are preserved, whereas coherences between subspaces of different $j$ and $j'$ are eventually fully lost. From this it is straightforward to see that the single-qubit coherences vanish for all four qubits, $\langle X_j \rangle = 0$. In contrast, the four-qubit operator $X_1 X_2 X_3 X_4$ has a non-zero expectation value, $\langle X_1 X_2 X_3 X_4 \rangle = 3/8$, which results in a lower bound 
\begin{align}
\bar{I}_{LB}=\frac{1}{4\cdot4\ln 2}\left[\langle X_1 X_2 X_3
    X_4 \rangle - \langle X_1\rangle \langle X_2\rangle \langle
    X_3\rangle \langle X_4\rangle \right]^2 =  0.0127.
\end{align}
\RefComm{For \textbf{Configuration 2} (qubits 1 and 2 encoded in Encoding A, qubits 3 and 4 encoded in Encoding B) the initial four-qubit state $\ket{\psi} = \ket{+}^{\otimes 4}$ will envolve into the density matrix $\rho = \rho_{12} \otimes \rho_{34}$ for long times. The state of the first and second, and third and fourth qubits, respectively, is given by
\begin{align}
\rho_{12} = \rho_{34} = \frac{1}{4} (\ket{00} \bra{00} + \ket{11} \bra{11} ) + \frac{1}{2} (|\Psi^+\rangle \langle\Psi^+|),
\end{align}
Therefore, all four single-qubit coherences vanish, $\langle X_j \rangle = 0$. However, due to the presence of the component which corresponds to the pair of Bell-states in the partially decohered four-qubit density matrix, the four-qubit operator $X_1 X_2 X_3 X_4$ has again a non-zero expectation value, $\langle X_1 X_2 X_3 X_4 \rangle = 1/4$, which in this case leads to 
\begin{align}
\bar{I}_{LB}=\frac{1}{4\cdot4\ln 2}\left[\langle X_1 X_2 X_3
    X_4 \rangle - \langle X_1\rangle \langle X_2\rangle \langle
    X_3\rangle \langle X_4\rangle \right]^2 = \frac{1}{256 \cdot \ln 2} = 0.0056 = 0.56 \%. 
\end{align}
Finally, for \textbf{Configuration 3}, with qubits 1, 2  and 3 encoded in Encoding A, and qubit 4 in Encoding B, one can show that the initial state $\ket{\psi} = \ket{+}^{\otimes 4}$ evolves for long enough times into
\begin{align}
\rho = \left[\frac{1}{8} \left( \ket{000}\bra{000} + \ket{111}\bra{111} \right) + \frac{3}{8} \left(\ket{\psi_1'} \bra{\psi_1'} + \ket{\psi_2'}\bra{\psi_2'} \right) \right]_{123} \otimes \frac{1}{2}\mathbb{1}_4
\end{align}
with the Dicke-type 3-qubit states
\begin{align}
\ket{\psi_1'} & = \frac{1}{\sqrt{3}} (\ket{100} + \ket{010} + \ket{001}),\nonumber\\ 
\ket{\psi_2'} & = \frac{1}{\sqrt{3}} (\ket{110} + \ket{101} + \ket{011}). \nonumber
\end{align}
Since for this state both the single-qubit coherences and $\langle X_1 X_2 X_3 X_4 \rangle$ vanish, one expects a vanishing spatial correlation measure, $\bar{I}_{LB} = 0$, in this limit.}\newline\\

\textit{2-qubit correlations - } Similarly, it is straightforward to obtain the expected behaviour for the long-time dynamics of two qubits undergoing perfectly correlated dephasing. In this case, the initial state $\ket{\psi} = \ket{+}^{\otimes 2}$ of a pair of qubits evolves for long times into
\begin{align}
\ket{\psi} \bra{\psi} \stackrel{t \rightarrow \infty}{\longrightarrow} \frac{1}{4} (\ket{00} \bra{00} + \ket{11} \bra{11} ) + \frac{1}{2} (|\Psi^+\rangle \langle\Psi^+|),
\end{align}
with the Bell state $\ket{\Psi^+} = \frac{1}{\sqrt{2}}(\ket{01} + \ket{10})$. Again, the single-qubit coherences vanish, $\langle X_1 \rangle = \langle X_2 \rangle = 0$, whereas the two-qubit operator $X_1 X_2$ saturates at $\langle X_1 X_2 \rangle = 1/2$, resulting in a lower bound for the two-qubit correlations of
\begin{align}
\bar{I}_{LB}=\frac{1}{4\cdot2\ln 2}\left[\langle X_1 X_2 \rangle - \langle X_1\rangle \langle X_2\rangle \right]^2 =  0.0451.
\end{align}